\documentclass[12pt]{article}
\usepackage{amsmath}
\usepackage{epsfig}
\usepackage{amssymb,amsfonts}

\advance\voffset by -2.0cm
\advance\hoffset by -1.25cm
\def\Im{\,{\rm Im}\, }

\newcommand{\bth}{\bar{\theta}}

\newcommand{\udpe}[1]{u^{(+,0)}_{#1}}
\newcommand{\udme}[1]{u^{(-,0)}_{#1}}
\newcommand{\udpz}[1]{u^{(0,+)}_{#1}}
\newcommand{\udmz}[1]{u^{(0,-)}_{#1}}
\newcommand{\udpme}[1]{u^{(\pm,0)}_{#1}}
\newcommand{\udmpe}[1]{u^{(\mp,0)}_{#1}}
\newcommand{\udpmz}[1]{u^{(0,\pm)}_{#1}}
\newcommand{\udmpz}[1]{u^{(0,\mp)}_{#1}}

\newcommand{\due}{du^{(\pm,0)}}
\newcommand{\duz}{du^{(0,\pm)}}

\newcommand{\uupe}[1]{u^{(+,0),#1}}
\newcommand{\uume}[1]{u^{(-,0),#1}}
\newcommand{\uupz}[1]{u^{(0,+),#1}}
\newcommand{\uumz}[1]{u^{(0,-),#1}}
\newcommand{\uupme}[1]{u^{(\pm,0),#1}}
\newcommand{\uupmz}[1]{u^{(0,\pm),#1}}

\newcommand{\fpp}{\varphi^{(+,+)}}
\newcommand{\fpm}{\varphi^{(+,-)}}
\newcommand{\fmp}{\varphi^{(-,+)}}
\newcommand{\fmm}{\varphi^{(-,-)}}

\newcommand{\psp}{\psi^{(+,0)}}
\newcommand{\psm}{\psi^{(-,0)}}
\newcommand{\chp}{\psi^{(0,+)}}
\newcommand{\chm}{\psi^{(0,-)}}

\newcommand{\tlpi}[1]{\theta^{#1(+,0)}}

\newcommand{\trpi}[1]{\theta^{#1(0,+)}}

\newcommand{\tlpid}[1]{\theta_{#1}^{(+,0)}}

\newcommand{\trpid}[1]{\theta_{#1}^{(0,+)}}

\newcommand{\dtl}{d^2\theta^{(+,0)}}
\newcommand{\hdtr}{d^2\theta^{(0,+)}}

\newcommand{\htrpi}[1]{\theta^{#1(0,+)}}

\newcommand{\htrpid}[1]{\theta_{#1}^{(0,+)}}

\newcommand{\N}{\mathcal{N}}

\def\be{\begin{equation}}
\def\ee{\end{equation}}
\def\ba{\begin{eqnarray}}
\def\ea{\end{eqnarray}}

\newcommand{\D}{{\cal D}}

\newcommand{\eg}{{\it e.g.~}}

\newcommand{\Q}{\Lambda_{\rm R}}

\newcommand{\bPhi}{{\phi}}
\newcommand{\GPhi}{{\Phi}}
\newcommand{\rmv}{\bar}

\newcommand{\cc}{\bar}

\newcommand{\om}{\omega_\lambda}
\newcommand{\thla}{\theta_\lambda}

\newcommand{\alo}{\alpha^{(1)}}
\newcommand{\alt}{\alpha^{(2)}}
\newcommand{\calo}{\cc \alpha^{(1)}}
\newcommand{\calt}{\cc \alpha^{(2)}}

\newcommand{\psio}{\psi^{(1)}}

\newcommand{\cpsio}{\cc\psi^{(1)}}

\def\be{\begin{equation}}
\def\ee{\end{equation}}
\def\bs{\begin{subequations}}
\def\es{\end{subequations}}

\newcommand\bea           {\begin{equation}\begin{array}l}
\newcommand\bearl         {\begin{array}{l}}
\newcommand\bearll        {\begin{array}{ll}}
\newcommand\eear          {\end{array}}

\def\Q{{\cal Q}}

\def\D{{\cal D}}
\newcommand{\beqa}{\begin{eqnarray}}
\newcommand{\eeqa}{\end{eqnarray}}
\newcommand{\beq}{\begin{equation}}
\newcommand{\eeq}{\end{equation}}

\title{\vspace{-1cm}\begin{flushright}{\small LMU -- ASC 09/09}\end{flushright}\vspace{2cm}\LARGE Obstructions and lines of marginal stability \\
from the world-sheet
\vspace*{0.5cm}}

\author{
Ilka Brunner$^{1,2}$\thanks{\tt E-mail: Ilka.Brunner@physik.uni-muenchen.de},
Matthias R.\ Gaberdiel$^{3}$\thanks{\tt E-mail: gaberdiel@itp.phys.ethz.ch}, 
Stefan Hohenegger$^{3}$\thanks{\tt E-mail: stefanh@itp.phys.ethz.ch}  
\vspace*{0.2cm} \\
and
Christoph A.\ Keller$^{4}$\thanks{\tt E-mail: ckeller@physics.harvard.edu} \\
\\ \\
${}^{1}${\small Arnold Sommerfeld Center, 
Ludwig Maximilians Universit\"at} \vspace*{-0.1cm} \\
{\small Theresienstr. 37, 
80333 M\"unchen, Germany} \vspace{0.3cm} \\
${}^{2}${\small Excellence Cluster Universe, 
Technische Universit\"at M\"unchen} \vspace*{-0.1cm} \\
{\small Boltzmannstr. 2, 
85748 Garching, Germany} \vspace{0.3cm} \\
${}^{3}${\small Institut f\"ur Theoretische Physik, 
ETH Z\"urich} \vspace*{-0.1cm} \\
{\small 8093 Z\"urich, Switzerland} \vspace{0.3cm} \\
${}^{4}${\small Jefferson Physical Laboratory, Harvard University}
\vspace*{-0.1cm} \\
{\small Cambridge, MA 02138, USA}
}
\date{\today}
\begin{document}
\maketitle

\begin{abstract}
The behaviour of supersymmetric D-branes under
deformations of the closed string background is studied using
world-sheet methods. We explain how
lines of marginal stability and obstructions arise from this point of
view.  We also show why $\N=2$ B-type branes may be 
obstructed against (cc) perturbations, but why such obstructions do 
not occur for $\N=4$ superconformal branes  at $c=6$, {\it i.e.}\  for 
half-supersymmetric D-branes on K3. Our analysis is based on 
a field theory approach in superspace, as well as  on
techniques from perturbed conformal field theory.
\end{abstract}

\newpage
\renewcommand{\theequation}{\arabic{section}.\arabic{equation}}


\section{Introduction}

The moduli space of supersymmetric D-branes  typically depends
on the closed string background of the setup, and 
its dimension and structure may change drastically as one varies the
background. One interesting effect that may happen is that a 
supersymmetric D-brane may cease to be  supersymmetric upon a small 
closed string perturbation that preserves the supersymmetry of
the closed string. In this case one usually says that there is 
an {\em obstruction} for adjusting the D-brane to the closed string 
deformation. Another even more drastic effect is that the D-brane may decay
into a superposition of D-branes; this is what happens as one crosses
a {\em line of marginal stability}. Both phenomena occur for 
supersymmetric D-branes in type II compactifications on Calabi-Yau manifolds. 
In this case, the combined bulk-boundary moduli space has a space-time
interpretation in terms of F-terms and D-terms of the low-energy theory.
F-term constraints are related to preserving an extended world-sheet supersymmetry 
(and hence to the absence of obstructions), whereas the requirement
of conformal invariance and charge integrality (that controls the lines of 
marginal stability) is encoded in the D-terms.

In this paper we study these phenomena from a world-sheet point of view
for theories with different amounts of supersymmetry. In particular, we 
shall concentrate on two cases: theories with $\N =1$ spacetime supersymmetry in 
four dimensions, in which case the corresponding world-sheet theory is an
$\N =2$ superconformal field theory \cite{Banks:1987cy},
and theories with $\N=1$ spacetime supersymmetry in six dimensions 
(which corresponds to $\N=2$ spacetime supersymmetry in four dimensions),  
in which case the world-sheet theory  has an $\N=4$ superconformal symmetry 
with $c=6$ \cite{Banks:1988yz}. In each case we shall give a world-sheet
interpretation of obstructions and lines of marginal stability. One of our
main results is that for D-branes that preserve the $\N=4$ superconformal 
symmetry with $c=6$ --- this is in particular the case for half-supersymmetric
branes on K3 --- obstructions cannot occur. 
\medskip

There are two natural and successful approaches to the analysis of
world-sheet theories. First one may regard them as 2-dimensional
supersymmetric field theories using superspace techniques, but without
using the conformal symmetry. In this formulation the supersymmetry
variation of the closed string perturbation is proportional to the
integral of a total derivative. On world-sheets without boundary these
perturbations thus preserve supersymmetry, but in the presence of
D-branes ({\it i.e.} world-sheets with boundaries) the total
derivatives may lead to non-trivial boundary terms.  Such non-trivial
boundary terms can sometimes be cancelled against the supersymmetry
variation of a correction term involving boundary fields
\cite{Warner:1995ay}. However, the boundary theory may not contain a
suitable boundary field, so that such a resolution is impossible. If
this is the case, the corresponding D-brane is obstructed against the
closed string perturbation.  As we shall explain, for world-sheet
theories with $\N=2$ supersymmetry, B-type branes may be obstructed in
this way against chiral perturbations, but not against twisted chiral
perturbations. 
On the other hand, there are no such obstructions for world-sheet
theories with $\N=4$ supersymmetry.  
\smallskip

This analysis can capture obstructions, but it does not detect
the potential presence of lines of marginal stability. In order to study 
the latter one also has to take into account the conformal symmetry; 
thus one is led to employ techniques of perturbed conformal field theory.  In
this approach deformations of the closed string background correspond
to perturbations of the conformal field theory by exactly marginal
bulk fields. In the situation with $\N=2$ superconformal world-sheet
symmetry, the corresponding bulk fields are either (cc), (ca), (ac) or
(aa) fields \cite{Dixon:1987bg,Distler:1988ms}. These fields are
always exactly marginal on world-sheets without boundary. In the
presence of boundaries, however, they may cease to be exactly
marginal. If this is the case they induce an RG flow on the boundary
that drives the brane to a configuration that is compatible with the
deformed closed string background \cite{Fredenhagen:2006dn}; the bulk
theory itself remains unaffected unless one also includes the
backreaction of the brane \cite{Keller:2007nd}. A typical example of
such an RG flow is what happens when one crosses a line of marginal
stability in the moduli space.

At a generic point of the moduli space, far away from lines of marginal
stability, no such RG flow is induced,
and thus the boundary condition remains conformal. Nevertheless the
bulk perturbation will generically modify the gluing conditions of the
symmetry algebra, including those that describe how the supercharges
are to be identified at the boundary. For $\N=2$ superconformal B-type
branes it may happen that a (cc) perturbation breaks the $\N=2$
superconformal symmetry at the boundary, {\it i.e.}\ that the brane is
obstructed.  
On the other hand, if the bulk and boundary theory are $\N=4$
superconformal with $c=6$, then we can show that the change in the
gluing condition for the $\N=2$ (or $\N=4$ generators) under a (cc)
perturbation can always be absorbed into a redefinition of the gluing
conditions that identify the left- and right-moving supercharges. Thus
$\N=4$ superconformal branes are not obstructed under such
perturbations.  \medskip

The paper is organised as follows. In section~2 we analyse the problem
from a field theory point of view. In particular, in
section~2.1 we use an $\N=2$ superspace formulation to explain
how obstructions may arise for $\N=(2,2)$ supersymmetric theories. 
Section~2.2 then deals with the case of $\N=(4,4)$
supersymmetry. The conformal field theory approach is described in
section~3. Section~3.2 and 3.3 deal with the case of $\N=2$ B-type
branes under (cc) and (ac) perturbations, respectively. The analysis
for the case with $\N=4$ superconformal symmetry is described in
section~3.4. We illustrate our findings in section~4 with two
examples, a simple brane configuration on a $T^4$ torus, as well as a
D-brane on K3 at the orbifold point $T^4/{\mathbb Z}_4$.  Finally,
section~5 contains our conclusions. There are two appendices where
some of the more technical material is described: appendix~A gives a
detailed account of our superspace conventions, in particular of the
harmonic superspace that is used in the $\N=(4,4)$ case. In appendix~B
we discuss constraints on the fusion rules of $\N=2$ and $\N=4$ chiral
primary operators.

\medskip


\section{The superspace analysis}\label{Sect:N2worldsheet}
In this section we analyse the problem of brane obstructions under 
deformations by bulk moduli, formulating the theory in terms
of an action in superspace. 
In a first step we will review the situation with
$\N=(2,2)$ supersymmetry, using an $\N=(2,2)$
superspace formulation, see \cite{Hori:2000ck,Hori}. 
In section~2.2 we shall then consider the situation with
$\N=(4,4)$ supersymmetry.

\subsection{The $\N=2$ analysis}\label{Sect:N2Analysis}

Let us begin by describing the two-dimensional world-sheet theory in standard 
$\N=(2,2)$ superspace
\begin{align}
\mathbb{R}^{(1,1|2,2)}=\mathbb{R}_L^{(1|2)}\times\mathbb{R}_R^{(1|2)}
=\{x^+,\theta^+,{\bth}^+\}\times \{x^-,\theta^-,{\bth}^-\}\ ,
\label{standardN2ss}
\end{align}
where the two factors represent the two light-cone sectors. Our conventions 
follow \cite{Hori}, and are described in appendix 
\ref{App:SuperspaceN2}. We parametrise an arbitrary supervariation as 
\be\label{ddef}
\delta   = \epsilon_+ \Q_-  - \epsilon_- \Q_+  
-\bar\epsilon_+ \bar\Q_-  + \bar\epsilon_- \bar\Q_+  \ .
\ee
Here the signs are chosen so that $\delta$ is a hermitian operator. In
order to study the supervariation in the presence of a boundary we  
need to specify which linear combinations of the supercharges are preserved 
at the boundary. We  shall always take the boundary to be along the 
line $x^+=x^-$, and we shall always consider B-type boundary conditions
\be
\epsilon \equiv \epsilon_+ = - \epsilon_- \ , \qquad
\bar\epsilon \equiv \bar\epsilon_+ = - \bar\epsilon_-\ .\label{BtypeCond}
\ee
For this boundary condition we shall then study 
chiral and twisted chiral deformations (that correspond to (cc) and
(ac) deformations, respectively). Note that this then covers already the 
general case since mirror symmetry exchanges A-type\footnote{In 
distinction to (\ref{BtypeCond}) A-type boundary conditions are given 
by $\epsilon\equiv\epsilon_+=\bar{\epsilon}_-$ and 
$\bar{\epsilon}\equiv\bar{\epsilon}_+=\epsilon_-$.} and B-type boundary 
conditions, as well as chiral and twisted chiral perturbations.

It will turn out that in both cases supervariations will only close up to
total derivatives, leading to boundary contributions which must be cancelled 
by introducing new terms involving fields on the boundary.
For twisted chiral perturbations this can always be achieved by using bulk fields
that are taken to the boundary. For chiral perturbations, on the other hand, we  
will in general need bona fide boundary fields that do not come
from the bulk. It is however not a priori clear that the boundary spectrum
contains the appropriate fields; if it does not, then we cannot cancel the boundary
contribution of the supersymmetry variation, and the brane will be obstructed.

\subsubsection{Chiral deformations}\label{Sect:ChirDeff}
Suppose now that we consider the variation of the (bulk) action by a
chiral superfield, {\it i.e.} by the term
\be\label{bper}
\Delta S = \int d^2x \, d\theta^- d\theta^+ \left. \Phi
\right|_{\bar\theta^\pm = 0} \ . 
\ee
{}From the conformal field theory point of view to be described below, this
corresponds to a (cc) deformation. 
If we write out the superfield  in components as in (\ref{chsup}), 
then $\Delta S$ equals 
\be
\Delta S = \int d^2x \, F(x^\pm) \ .\label{bpercom} 
\ee
We are interested in the supersymmetry variation described by $\delta  $
of this perturbation. To this end we calculate
\be
\delta  \, \Delta S = \int d^2x \, d\theta^- d\theta^+ 
\left(\epsilon_+ \Q_-  - \epsilon_- \Q_+  
-\bar\epsilon_+ \bar\Q_-  + \bar\epsilon_- \bar\Q_+  \right) \Phi \ .
\ee
It is easy to see that the terms involving $\Q_\pm $ act trivially. To
evaluate the other two terms we use that $\bar\D_\pm \Phi=0$, and thus
find
\begin{eqnarray}
\delta  \, \Delta S & = & 
\int d^2x \, d\theta^- d\theta^+ \left( 
2 i \bar\epsilon_+ \theta^- \partial_- \Phi - 2 i 
\bar\epsilon_- \theta^+ \partial_+ \Phi \right) \nonumber \\
& = & 
- 2 i \int d^2x \, \left( \bar\epsilon_- \partial_+ \psi_-
+ \bar\epsilon_+ \partial_- \psi_+ \right) \ . 
\end{eqnarray}
The last term is a total derivative and thus vanishes on a
world-sheet without boundary. In the presence of a boundary along 
$x^1=0$, the $\partial_1$ derivative gives the contribution
\begin{eqnarray}
\delta   \Delta S = -i \int_{x^1=0} dx^0 \, 
\left( \bar\epsilon_- \psi_- - \bar\epsilon_+ \psi_+ \right) \ . 
\end{eqnarray}
The full perturbation we are interested in contains also the complex
conjugate of (\ref{bper}), namely
\be
\overline{\Delta S} =  \int d^2x \, d\bar\theta^+ 
d\bar\theta^- \left. \bar\Phi \right|_{\theta^\pm = 0} \ ,
\label{aasuperspacedef}
\ee
where $\bar\Phi$ is the complex conjugate of $\Phi$, and thus defines
an anti-chiral superfield. By a similar calculation to the above 
(using the expansion of the anti-chiral superfield $\bar\Phi$ as in 
(\ref{achsup})) one finds that 
\begin{eqnarray}
\delta   \overline{\Delta S} & = & 2 i 
\int d^2x \, d\bar\theta^+ d\bar\theta^- \left(
\epsilon_+ \bar\theta^- \bar\theta^+ \partial_- \bar\psi_+ 
- \epsilon_- \bar\theta^+ \bar\theta^- \partial_+ \bar\psi_- \right) 
\nonumber \\
& = & 
i \int_{x^1=0} dx^0 \, \left(\epsilon_- \bar\psi_- - \epsilon_+
\bar\psi_+ \right) \ . 
\end{eqnarray}
Altogether the supersymmetry variation of the total
chiral deformation equals
\be\label{total}
\delta   (\Delta S + \overline{\Delta S} ) = 
i \int_{x^1=0} dx^0 \, \left(\bar\epsilon_+ \psi_+ 
-\bar\epsilon_- \psi_- + 
\epsilon_- \bar\psi_- - \epsilon_+
\bar\psi_+ \right) \ . 
\ee
Using the explicit form of the B-type boundary conditions (\ref{BtypeCond})
this can be rewritten as 
\begin{eqnarray}\label{totalB}
\delta   (\Delta S + \overline{\Delta S} ) & =  & 
i \int_{x^1=0} dx^0 \, \left(
-\epsilon (\bar\psi_+ + \bar\psi_-) 
+ \bar\epsilon (\psi_+ +\psi_-) \right) \nonumber \\
& = & 
-i \int_{x^1=0} dx^0d\bar\theta \,
\epsilon \, \bar\Phi 
+ i \int_{x^1=0} dx^0 d\theta\,  \bar\epsilon \, \Phi  \ ,
\end{eqnarray}
where we used the B-type conditions to define the boundary Grassmann variables
$\theta = \theta^+ + \theta^-$ and $\bar \theta = \bar \theta^+ +\bar \theta^-$.
In general, this term cannot be cancelled by a boundary term involving
only bulk fields. Thus B-type branes may be obstructed under (cc)
deformations. From a space-time point of view, this reflects the fact that the
(cc) fields enter the superpotential $W$, and allowed motions in parameter space
are constrained to the zero locus of $\partial W$. Note also that the 
(cc) fields as well as the B-type boundary conditions survive the topological B-twist; 
thus the (holomorphic) superpotential $W$ is calculable in the topological B-model.

Sometimes it is possible to preserve the 
full $\N=2$ by introducing additional boundary degrees of freedom with 
appropriate supersymmetry transformations. This has, in particular, been done
in the context of Landau Ginzburg models
\cite{Kapustin:2002bi,Brunner:2003dc} (for a review see 
\cite{Hori:2004zd,Jockers:2007ng}). 
In some situations, however, even this is not enough
to preserve $\N=2$ supersymmetry, as can be seen for example in
\cite{Ashok:2004xq,Hori:2004ja}.  
While the $\N=2$ supersymmetry may thus be broken by the deformation, it is
always possible to cancel the variation of the $\N=1$
supersymmetry variation that corresponds to $\epsilon = \bar\epsilon$. 
In fact, since the $\N=1$ supersymmetry is a gauge symmetry in string theory, 
this is crucial for the consistency of the perturbation. The boundary term that 
cancels the $\N=1$ variation is simply given by
\be\label{Bdef1}
{\cal B} = - i \int_{x^1=0} dx^0 (\phi - \bar\phi) \ ,
\ee
whose variation yields
\be
\delta   {\cal B} = -
i \int_{x^1=0} dx^0 \, 
\left( \epsilon (\psi_+ + \psi_-) - 
\bar\epsilon (\bar\psi_+ + \bar\psi_-) \right) \ .
\ee
This indeed cancels (\ref{totalB}) for $\epsilon = \bar\epsilon$.

\subsubsection{Twisted chiral deformations}\label{Sect:TChirDeff}

The other possibility is that we modify the action by 
a twisted chiral perturbation (corresponding to an (ac) field)
\be\label{twbper}
\Delta S = \int d^2x \, d\bar\theta^- d\theta^+ \left. U
\right|_{\bar\theta^+=\theta^- = 0} \ ,
\ee
where $U$ is a twisted chiral supermultiplet. Since the calculation of the
supersymmetry variation proceeds much like in the case of  the
twisted perturbation, we will omit the details. The resulting
boundary term is
\be
\delta  \Delta S = i \int_{x^1=0} dx^0 \left(\epsilon_+ \chi_+
-\bar\epsilon_- \bar\chi_- \right) \ . 
\ee
Combining this with its complex conjugate, one obtains
\be\label{218}
\delta  (\Delta S +\overline{\Delta S})= 
i \int_{x^1=0} dx^0 \left(\epsilon_+\chi_+
-\bar\epsilon_-\bar\chi_-  
-\bar\epsilon_+ \bar\chi_+ + \epsilon_- \chi_-\right) \ .
\ee
This boundary term can always be cancelled by adding the boundary
integral
\be\label{Btwdef1}
{\cal B} = - i \int_{x^1=0} dx^0 (v - \bar{v}) \ ,
\ee
where $v$ is the lowest component of the twisted chiral superfield,
while its complex conjugate $\bar{v}$ is the lowest component of the
twisted anti-chiral superfield (see appendix \ref{App:SuperspaceN2}). 
Indeed, the variation of (\ref{Btwdef1})
cancels precisely (\ref{218}). This ties in with the expectation that
B-type branes are never obstructed against (ac) deformations 
\cite{Brunner:1999jq}. The complete story is however more complicated 
since there are in general lines of marginal stability in the (ac) moduli space 
along which D-branes will decay 
\cite{Douglas:2000ah,Douglas:2000gi,Walcher:2004tx}. We will return to 
this phenomenon in section~3.

\subsection{Theories with $\N=4$}

We now wish to generalise the results of the previous section to cases where 
$\N=(2,2)$ supersymmetry is enhanced to $\N=(4,4)$. We shall discuss the
problem using an $\N=4$ superspace formulation. Different $\N=4$ superspace
formulations are known, in particular
projective superspace 
\cite{Gates:1984nk,Rocek:1991vk,Buscher:1987uw,Lindstrom:1994mw}, 
harmonic superspace \cite{Ivanov:1994er,Ivanov:2004re}, as well as 
conventional $\N=(4,4)$ superspace using constrained superfields 
\cite{Gorovoi:1991td}. In the following we shall employ 
harmonic superspace; some technical details 
can be found in appendix \ref{App:Superspace4} (see also 
\cite{Galperin:1984av,Galperin:1984bu,Hartwell:1994rp,Galperin:2001uw}).

\subsubsection{$\N=4$ harmonic superspace}
The simplest way to enhance the superspace (\ref{standardN2ss}) to
$\N=(4,4)$ is to promote all Grassmann coordinates to doublets under
an $SU(2)\times SU(2)$ R-symmetry group\footnote{In the following we
shall drop the superscripts $\pm$ of the Grassmann variables in order
to avoid cluttering our formulae. Note that the Grassmann variables
also transform as doublets with respect to another $SU(2)$. We shall
call the former $SU(2)$ action (which acts on the indices $i$ and
$\bar{\imath}$) `flavour' $SU(2)_{f}$, while the latter $SU(2)$ (rotating
the indices $a$ and $\bar{a}$) will be referred to as `colour'
$SU(2)_{c}$.  This notation follows the nomenclature of
\cite{Berkovits:1994vy}.}  
\be
\mathbb{R}^{(1,1|4,4)}=\mathbb{R}_L^{(1|4)}\times
\mathbb{R}_R^{(1|4)}=\{x^+,\theta_i^a\}\times 
\{x^-,\theta_{\bar{\imath}}^{\bar{a}}\}\ ,\label{standardss} \ee where
$i,\bar{\imath}=1,2$ and the index $a$ and $\bar{a}$ distinguishes
spinors from their conjugates (see (\ref{DoubletGrassmann})).  The
explicit construction of 
(off-shell) superfields on $\mathbb{R}^{(1,1|4,4)}$ is a rather
delicate issue (see {\it e.g.}\ \cite{Gorovoi:1991td}), and we
therefore prefer to formulate our theory on a slightly modified
(harmonic) superspace. To this end we define \be
\mathbb{H}^{(1,1+4|4,4)}=\mathbb{R}^{(1,1|4,4)}\times
\frac{SU(2)_L}{U(1)_L}\times \frac{SU(2)_R}{U(1)_R}
=\{x^+,\theta^{a(\pm,0)},\udpme{i}\}\times
\{x^-,\theta^{\bar{a}(0,\pm)}, \udpmz{\bar{\imath}}\}\ , \label{harmss} \ee
where we have introduced new variables
\begin{align}
&\{\udpme{i}\}\in \frac{SU(2)_L}{U(1)_L} &&\text{and} &&
\{\udpmz{\bar{\imath}}\}\in \frac{SU(2)_R}{U(1)_R}\ .\label{trueharmonics}
\end{align}
The $SU(2)$ properties of these variables become apparent if one
writes them in matrix form like in (\ref{A25}). The essential idea is
now to identify these groups with the left- and right-moving $SU(2)_f$,
respectively.\footnote{The  
actual construction of $\mathbb{H}^{(1,1+4|4,4)}$ as a coset space is
a little bit more  
involved and will not be presented here since it is only of minor
importance for the rest  
of this work. A detailed description of the construction can however,
for example,  
be found in \cite{Galperin:2001uw} (see also
\cite{Antoniadis:2007cw}).} This allows us to  
define new Grassmann variables via projection with the harmonic variables
\begin{align}
&\theta^{a(\pm,0)} =\theta_i^a\uupme{i}, &&\text{and} &&
\theta^{\bar{a}(0,\pm)}=\bar{\theta}_{\bar{\imath}}^{\bar{a}}
\uupmz{\bar{\imath}} \,,
\end{align}
(for more details see appendix \ref{App:SuperspaceHS}). In these
conventions, the analogue of the supervariation  (\ref{ddef}) takes the form
\be
\delta=\epsilon^{a(-,0)}\Q_a^{(+,0)}+\epsilon^{a(+,0)}\Q_a^{(-,0)}
+\epsilon^{\bar{a}(0,-)}\Q_{\bar{a}}^{(0,+)}
+\epsilon^{\bar{a}(0,+)}\Q_{\bar{a}}^{(0,-)}\ .
\label{deltaN4}
\ee
\smallskip

We also need to formulate the $\N=4$ analogue of the chiral and
twisted-chiral deformations of the $\N=2$ theory which we have
discussed in section~2.2.  In the $\N=4$ language, both deformations
can be described in a unified manner as 
(see \cite{Ivanov:1994er})
\begin{align}
S_{\text{def}}=\int d^2x\int \due\int\duz \int \dtl\int 
\hdtr  \tlpi{a}\htrpi{\bar{a}}
\zeta_{a\bar{a}}\Phi^{(+,+)} \ ,\label{streamlineMASSDEF}
\end{align}
where $\zeta_{a\bar{a}}$ is a parameter which essentially picks out
various components of the $\N=4$ superfield. Notice that upon focusing
on a specific $\N=(2,2)$ subalgebra, these components can be precisely
identified with the (cc), (ac), (ca) and (aa) perturbations which we
have discussed in section \ref{Sect:N2Analysis}. Note also that the
integrand of this superspace integral contains explicitly the
Grassmann variables, which naively indicates that the deformation does
not preserve supersymmetry (even without boundaries). However, given
the special properties of the superfields $\Phi^{(+,+)}$ and the fact
that $\zeta_{a\bar{a}}$ is a $u$-independent constant, the above terms
are actually supersymmetric. One way to see this is to rewrite
(\ref{streamlineMASSDEF}) in component language.  We first perform the
Grassmann integration in (\ref{streamlineMASSDEF}) which yields
\begin{align}
S_{\text{def}}&=\int d^2x\int \due\int\duz \int \dtl\int \hdtr  
\tlpi{a}\htrpi{\bar{a}} \, \zeta_{a\bar{a}}\, \Phi^{(+,+)}\nonumber\\
&=\int d^2x\int \due\int\duz \int d\tlpid{a}\int d\htrpid{\bar{a}}  \,
\zeta_{a\bar{a}}\, {\Phi^{(+,+)}}_{\big|\tlpi{a}=\htrpi{\bar{a}}=0}\nonumber\\
&=\int d^2x\int \due\int\duz \, \zeta_{a\bar{a}} \, F^{a\bar{a}}.
\end{align}
Since the integrand is independent of the harmonic variables the 
$\int \due$ and $\int\duz$ integrals are trivial, and the deformation
can be written as 
\begin{align}
S_{\text{def}}&=\int d^2x\, \zeta_{a\bar{a}}\, F^{a\bar{a}}\ .\label{DEFcomponent}
\end{align}
In particular, this term has the same structure as the $\N=2$ terms 
(\ref{bpercom}) or (\ref{twbper}). Since the supervariaton of $F^{a\bar{a}}$ 
is a total derivative
\begin{align}
\delta F^{a\bar{a}}=2i\, \epsilon^a_i \, \partial_+ \psi^{i\bar{a}}
+2i\, \epsilon^{\bar{a}}_{\bar{\imath}} \, \partial_- \psi^{\bar{\imath}a}\,,
\end{align}
it is now obvious that (\ref{DEFcomponent}) is in fact invariant under 
supersymmetry transformations (at least as long as there are no boundaries).

\subsubsection{Boundary conditions and deformations}

Before we can study the behaviour of boundary conditions under these bulk
deformations we need to review the structure of the $\N=4$ preserving
boundary conditions. The automorphism (R-symmetry) group of the $\N=4$ 
algebra is $SU(2)_{c}\times SU(2)_{f}\cong SO(4)$. Thus a general $\N=4$ 
preserving boundary condition will be labelled by an element in this
group, where  
the corresponding automorphism describes how the left- and right-moving
supercharges are identified at the boundary $x^+=x^-$. 

In the context of string theory, not all such boundary conditions are
of interest.  
This comes from the fact that the overall $\N=(1,1)$ supersymmetry is 
a gauge symmetry in string theory, and thus that the corresponding
supercharges  
must be identified without any non-trivial automorphism
\cite{Ooguri:1996ck}. 
Since we want to impose a fixed $\N=1$ symmetry, we
restrict ourselves to identifications of the form
\be\label{2.29}
\Q^i_a = U^i{}_{\bar\jmath} \, \Q^{\bar\jmath}_a \ ,
\ee
where $U^i{}_{\bar\jmath}$ is a given $SU(2)$ matrix. This corresponds to
imposing the boundary conditions for the Grassmann variables 
\be\label{thetarel}
\theta_i^{a} = U_i{}^{\bar\jmath} \, \theta_{\bar\jmath}^{a} \ ,
\ee
where $U_i{}^{\bar\jmath}$ is the inverse of the matrix in (\ref{2.29}). We
should also mention that from the harmonic point of view different
choices of the matrix ${U_i}^{\bar{\jmath}}$ correspond to different
parametrisations of the coset space (\ref{trueharmonics}), and
therefore can also be interpreted as simple coordinate transformations
in our fully covariant notation.  

In the presence of such a boundary condition we are then interested in the 
supervariations for which
\be
\epsilon_i^a = U_i{}^{\bar\jmath} \, \epsilon_{\bar\jmath}^a \ ,
\ee
again with the same $U_i{}^{\bar\jmath}$. If we now perturb the theory by 
the bulk deformation (\ref{streamlineMASSDEF}), then the supervariation 
equals, using the component language (\ref{DEFcomponent})
\begin{align}
\delta S_{\text{def}}&=2i\int d^2x\, \zeta_{a\bar{a}}
\left(\epsilon^a_i\partial_+
  \psi^{i\bar{a}}+\epsilon^{\bar{a}}_{\bar{\imath}}
\partial_- \psi^{\bar{\imath}a}\right)\nonumber\\
&=2i\int dx^0\zeta_{a\bar{a}}\left(\epsilon^a_i 
\psi^{i\bar{a}}-\epsilon^{\bar{a}}_{\bar{\imath}}\psi^{\bar{\imath}a}\right)
=2i\int dx^0\, \zeta_{[a\bar{a}]}\epsilon^a_i \psi^{i\bar{a}}\ .
\label{componentoffence}
\end{align}
In the last step we have used that (\ref{thetarel}) implies that also 
$\psi^{ia}$ and 
$\psi^{\bar\imath a}$ satisfy a similar relation at the boundary.
Since $\zeta_{[a\bar{a}]}$ is an antisymmetric $2\times 2$ matrix, we can 
always write it as
\begin{align}
\zeta_{[a\bar{a}]}=\epsilon_{a\bar{a}}
\left(\epsilon^{b\bar{b}}\zeta_{b\bar{b}}\right)\equiv 
\epsilon_{a\bar{a}}g\ .
\end{align}
It is then easy to see that the above supervariation is precisely cancelled
by the supervariaton of the boundary term
\begin{align}
S_{\text{def}}^{\text{bdy}}=\int dx^0\, g\,  \varphi^{ij}\, \epsilon_{ij}\ .
\end{align}
This implies that on the level of the superspace analysis, 
$\N=4$ supersymmetric branes are never obstructed.

\section{Conformal field theory analysis}\label{s:CFT}
\setcounter{equation}{0}
In the previous section we showed, using a superspace
approach on the world sheet
that $\N=4$ supersymmetric branes are not obstructed against
bulk perturbations that preserve the $\N=(4,4)$ supersymmetry in the bulk.
Note that this did not deal with the question of the conformal symmetry
of the theory, which is of course crucial \eg in applications
to string theory. 
In this section we thus address the same problem using conformal
field theory methods. Again we begin by studying the analysis from an 
$\N=2$ point of view. 

\subsection{Generalities for $\N=2$ theories}

We are interested in (closed string) perturbations that preserve the 
$\N=(2,2)$ superconformal symmetry in the bulk.\footnote{In the 
context of an $\N=(4,4)$ theory, these perturbations will then also
preserve the $\N=(4,4)$ symmetry in the bulk \cite{Seiberg:1988pf}.} 
There are four different types of perturbations that have this property,
and they are usually referred to as (cc), (aa), (ac) and (ca).
In all cases, the perturbation is of the form
\begin{align}
&\int d^2z\,  \widehat\Phi(z,\bar{z})= 
\int d^2z  \bigl( G_{-1/2} \tilde{G}_{-1/2} \Phi\bigr)(z,\bar{z})\ ,
&&\text{where}&&\widehat\Phi = G_{-1/2} \tilde{G}_{-1/2} \Phi \ ,\label{CFTdef}
\end{align}
and $G_r$, $\tilde{G}_r$ are the modes of the left- and right-moving
$\N=1$ supercurrent, which in terms of the supercurrents of the
$\N=2$ algebra is simply $G_r=G^+_r + G^-_r$  and similarly for the
right-movers (see appendix~\ref{app:N4}.)\footnote{Unlike in
section~2, $\pm$ now denotes the U(1) charge of the operators,
whereas right-movers are denoted by a tilde. Bar will be used later
on for complex conjugation.}  
Depending on which case one considers 
$\Phi$ in (\ref{CFTdef}) is a chiral 
(c) or anti-chiral (a) primary with respect to the left- and
right-moving $\N=2$ algebra.  
So for example, in the (cc) case, we have 
\be
G^+_{-1/2} \Phi = \tilde{G}^+_{-1/2} \Phi = 0 \ , \qquad
\hbox{so that} \qquad
\widehat\Phi = G^-_{-1/2} \tilde{G}^-_{-1/2} \Phi \ . 
\ee
This is the perturbation that corresponds to the `chiral deformation'
of section~2.1.1.  
As is well known \cite{Lerche:1989uy}, for chiral primaries the U(1) charge 
and the conformal dimension is related as $h=q/2$, while for
anti-chiral primaries  
we have instead $h=-q/2$. For the marginal bulk perturbations we need that
$h=\bar{h}=\tfrac{1}{2}$, and thus $q=\pm 1 = \bar{q}$. In particular, 
$\widehat{\Phi}$ then always has vanishing U(1) charges. 
\smallskip

We are interested in such perturbations in the presence of a
boundary. As already mentioned before, there are two natural boundary
conditions for $\N=(2,2)$ theories, namely A-type and B-type
branes. In terms of the conformal field theory description, they are
characterised by their gluing conditions, specifying how the left- and
right-moving fields are identified at the boundary. If we take the
boundary to be the real axis, the two cases are
\begin{align}
T(z) = \tilde T(\bar z)\ , && G^\pm(z) = \tilde{G}^\mp(\bar z)\ , && J(z) =
-\tilde J(\bar z)\ , && \textrm{for}\ z = \bar z && \textrm{(A-type)}\
, \label{Atype}\\
T(z) = \tilde T(\bar z)\ , && G^\pm(z) = \tilde{G}^\pm(\bar z)\ , && J(z) =
\tilde J(\bar z)\ , && \textrm{for}\ z = \bar z && \textrm{(B-type)}\ ,
\label{Btype}
\end{align}
where the right-moving fields are denoted by a tilde. 
Because of mirror symmetry the behaviour of A-type D-branes under
(cc) and (ac) deformations is the same as that of B-type D-branes under
(ac) and (cc) deformations, respectively, and similarly for (ca) and (aa)
perturbations. It is therefore sufficient to concentrate on the case of B-type 
boundary conditions. Furthermore, the analysis of (aa) deformations is
essentially identical to that of a (cc) deformation, and similarly for 
(ca) and (ac). Thus we shall concentrate on B-type branes under
(cc) and (ac) deformations.

In either case, there are two issues to consider. First, as discussed in 
\cite{Fredenhagen:2006dn} (see also 
\cite{Fredenhagen:2007rx,Gaberdiel:2008fn}), 
a bulk perturbation may break the conformal symmetry on the boundary 
and induce a non-trivial RG flow. This will be the case provided that the
bulk perturbation $\widehat\Phi$ will switch on a marginal 
boundary field as it approaches the boundary.  Then the brane will
generically flow a finite distance
to a different boundary condition, and we should not
expect to be able to say much about the symmetries it preserves.

On the other hand, if no such marginal or relevant operator is
switched on, then the brane will only adjust infinitesimally to an
infinitesimal bulk perturbation.  Then we can ask, following
\cite{Fredenhagen:2007rx}, to which extent the gluing condition will
be modified.

In the following we shall analyse these questions separately for
the (cc) and (ac) perturbations.

\subsection{$\N=2$ B-type branes under (cc) deformations}

Let us first ask whether the bulk perturbation by the (cc) field
$\widehat\Phi$ will break the conformal invariance of the 
boundary theory. To this end we need to consider the 
bulk-boundary OPE as the field $\widehat\Phi$ approaches
the boundary. By the usual doubling trick \cite{Cardy:1984bb}
we can replace the bulk field $\widehat\Phi$ by two chiral fields, one
in the upper half-plane, and one in the lower half-plane. Since
we are dealing with a B-type boundary condition, both of these fields
are $G^-_{-1/2}$ descendants of chiral primary fields with $q=1$ and 
$h=\tfrac{1}{2}$, and we shall denote them by $\phi_c$ and
$\bar{\phi}_c$, respectively.

The possible boundary fields that are switched on are constrained
by the fusion rules of the $\N=2$ superconformal algebra
\cite{Mussardo:1988av,Gaberdiel:1993mt}. In particular, it follows from 
the analysis of appendix~\ref{App:FusionN2} that the fusion of two chiral primary
fields with $h=\tfrac{1}{2}$ and $q=1$ only contains the
superconformal families 
\be \label{ccfusion}
\phi_c \otimes \bar{\phi}_c = [\varphi_{h=1,q=2}] 
\oplus [G^+_{-1/2} \varphi^+_{q=1}] \ .
\ee
The first term is the usual chiral ring product and represents the so-called 
`even fusion rules'; the second term describes the odd fusion that is possible
in this case. Only the U(1) charge of the fields $\varphi^+$ is fixed
to be $q=1$, but  
the conformal dimension of $\varphi^+$ is not constrained; in particular
there may therefore be more than one such channel. However, for each
of these fields 
the usual unitarity bound implies that $h\geq \tfrac{1}{2}$. Furthermore,
$h=\tfrac{1}{2}$ is excluded, since $\varphi^+$ would then be a
chiral primary field for which $G^+_{-1/2} \varphi^+=0$.

Since fusion rules describe the product structure of the corresponding
superconformal families, we can then also conclude what the leading
terms in the OPE of $G^-_{-1/2}\phi_c$ must be. In fact, 
from (\ref{ccfusion}) and U(1) charge conservation it follows that 
\be\label{N2fusion}
(G^-_{-1/2} \phi_c) \otimes (G^-_{-1/2} \bar{\phi}_c) = 
G^-_{-1/2} G^-_{-3/2} \varphi_{h=1,q=2} + 
G^-_{-1/2} \varphi^+_{q=1} + \hbox{higher descendants .}
\ee
Because the conformal dimension of $\varphi^+$ satisfies
$h>\tfrac{1}{2}$, it then follows that none of the fields on the right hand
side are marginal or relevant. This then implies that no marginal
or relevant boundary field is switched on by the perturbation by
$\widehat\Phi$. Thus the perturbation by the (cc) field $\widehat\Phi$
does not break the conformal symmetry. 

\subsubsection{Modifying the gluing condition}

Since the bulk perturbation does not break the conformal invariance of the 
boundary condition, we can ask whether it will affect the
superconformal gluing  
conditions, in particular those corresponding to $G^\pm$. 
To analyse this effect, we need to apply the analysis of 
\cite{Fredenhagen:2007rx} to the present context. To first order in
the bulk perturbation, the  
change in the gluing condition for $G^\pm$ is determined by 
(compare eq.\ (3.9) of \cite{Fredenhagen:2007rx} --- we are using 
here that the automorphism $\omega$ is trivial for B-type branes)
\be\label{1}
\Delta G^\pm = \lambda  \lim_{y\rightarrow 0}
\int_{\mathbb{H}_+} d^2 w \, \bigl(G^\pm(z) - \tilde{G}^\pm(\bar z)\bigr) \,
\widehat\Phi(w,\bar w)\ ,
\ee 
in the limit where $z$ approaches the boundary, {\it i.e.}\ 
$y=\Im z\rightarrow 0$. Using the doubling trick  \cite{Cardy:1984bb}
we can think of this as a chiral correlator on the full plane, where
$\tilde{G}^\pm(\bar z)$ is the usual chiral $G^\pm$ field at the image
point, and we write $\widehat\Phi$ as a product of two chiral fields, 
namely $G^-_{-1/2}\phi_c$ at $w$, and $G^-_{-1/2}\bar\phi_c$ at $\bar{w}$.
Then the above expression has four poles. For the case of the field 
$G^+$ each of them is of the form
\begin{eqnarray} \label{glueG}
G^+(z) \, \bigl( G^-_{-1/2}\phi_c\bigr)(w)  & \simeq & 
\frac{1}{(z-w)^2} V(G^+_{1/2} G^-_{-1/2} \phi_c,w) + 
\frac{1}{(z-w)}V(G^+_{-1/2} G^-_{-1/2}\phi_c,w) \nonumber \\
& = & \frac{2}{(z-w)^2} \, V(\phi_c,w) + 
\frac{2}{(z-w)} \, \partial_w V(\phi_c,w)  \nonumber \\
&= & 2 \, \frac{d}{dw}
\Bigl( \frac{1}{z-w} \, V(\phi_c,w) \Bigr) \ .
\end{eqnarray}
Here we are using the same notation as in \cite{Fredenhagen:2007rx},
and $V(\phi,z)$ denotes the field corresponding to the state $\phi$ at the 
position $z$. On the other hand, for $G^-$ there is no pole at all. 
Thus the gluing condition for $G^-$ will not be modified. For 
$G^+$, on the other hand, using the fact that (\ref{glueG}) is 
a total derivative, we can do the integral in (\ref{1}) and find
\be\label{2}
\Delta G^+ = 2 \lambda \pi\, \lim_{y\rightarrow 0}
\Bigl[  \phi_c(z) \, \bigl(G^-_{-1/2}\bar{\phi}_c\bigr)(\bar{z})
- \bigl(G^-_{-1/2} \phi_c \bigr)(z) \, \bar{\phi}_c(\bar{z}) \Bigr] \ .
\ee
In general this limit is non-trivial since the fusion rules 
(\ref{N2fusion}) allow for non-trivial terms.
In fact, since $\phi_c$ is a fermionic field, the linear combination 
that appears in (\ref{2}) is just the $G^-_{-1/2}$ descendant
of (\ref{ccfusion}), and hence we have that 
\be
\Bigl[  \phi_c(z) \, \bigl(G^-_{-1/2}\bar{\phi}_c\bigr)(\bar{z})
- \bigl(G^-_{-1/2} \phi_c \bigr)(z) \, \bar{\phi}_c(\bar{z}) \Bigr] 
= G^{-}_{-1/2} \,\varphi_{h=1,q=2} \ + \ 
G^-_{-1/2} G^+_{-1/2} \varphi^+_{q=1} \ + 
\ \hbox{desc.} 
\ee
The limit $y\rightarrow 0$ is then well defined, and only the first
term survives, thus giving 
\be\label{Gpch}
\Delta G^+ =  2\lambda \pi\, G^{-}_{-1/2} \,\varphi_{h=1,q=2} \ .
\ee
Hence the gluing condition for $G^+$ is  modified unless
$\varphi_{h=1,q=2}$ vanishes, 
whereas that for $G^-$ is unmodified. Because of unitarity we also need 
to perturb by the conjugate (aa) field, which in turn only changes
$G^-$, but not $G^+$. If the boundary condition just preserves
the $\N=2$ superconformal algebra, its boundary spectrum does 
not contain a chiral field of conformal dimension $h=1$ and U(1) charge
$q=2$. Thus the correction term 
$\Delta G^+ = G^-_{-1/2} \varphi_{h=1,q=2}$ is not part of the symmetry 
algebra of the boundary theory, and hence this modification cannot
be absorbed into a redefinition of the gluing condition. In this case
the bulk perturbation therefore breaks the $\N=2$ superconformal
symmetry of the boundary condition. This means that the brane is 
obstructed against this (cc) deformation.

In the $\N=2$ case, the condition for whether there is an obstruction
is therefore whether the bulk boundary correlator \be\label{bbco}
\langle \Phi_{cc} (z,\bar{z}) \, \varphi_{h=1,q=2}(x) \rangle \ee
vanishes or not. In fact, this correlator is a `topological' quantity,
which can be calculated in topological string theory. For the case of
a 3-dimensional Calabi-Yau manifold, it is (after spectral flow)
precisely equal to the bulk boundary correlator that was calculated in
\cite{Baumgartl:2007an} using the Kapustin-Li formula
\cite{Kapustin:2003ga}. As was explained in \cite{Fredenhagen:2007rx},
this correlator vanishes if and only if the bulk field is BRST-exact
when brought to the boundary, which in turn is the condition of
\cite{Hori:2004ja} for the absence of an obstruction to first
order. In the case of K3, spectral flow works slightly differently,
and (\ref{bbco}) becomes precisely the `charge' of the bulk
field. This controls the obstruction theory for K3, as was explained
in \cite{Brunner:2006tc}.

\subsubsection{Restoring the $\N=1$ superconformal symmetry}

The above analysis shows that the $\N=2$ superconformal symmetry is
broken by a (cc) perturbation if (\ref{bbco}) does not vanish. If this is the 
case, also the $\N=1$ superconformal symmetry is broken. From a string
theory point of view this is {\em not acceptable} since the $\N=1$
superconformal  
symmetry is a gauge symmetry. In order to explain how we can restore this
symmetry we need to consider a real perturbation, {\it i.e.}\ 
we need to add to the (cc) perturbation its hermitian conjugate, which is
then an (aa) perturbation. The above considerations imply that 
the gluing map for $G=G^+ + G^-$ changes by 
\be
\Delta G = 2 \lambda \pi\, 
\Bigl[ G^-_{-1/2} \varphi_{h=1,q=2} +
G^+_{-1/2} \varphi_{h=1,q=-2} \Bigr] \ , \label{bulkglueshift}
\ee
where $\varphi_{h=1,q=-2}$ is the field that appears in the even fusion rule of 
$\phi_a$ with $\bar{\phi}_a$. 
To restore the original gluing condition, we now turn on the 
boundary perturbation 
\be
\Delta S_{bd} = i \lambda \int dx\, 
\Bigl[  \varphi_{h=1,q=2} + \varphi_{h=1,q=-2}  \Bigr] \ . \label{Sbd}
\ee
The effect of boundary perturbations on the gluing map were 
analysed in \cite{Recknagel:1998ih}. The relevant term comes 
from the first order pole as the chiral field $G$ approaches the 
boundary field in (\ref{Sbd}). Since $\varphi_{h=1,q=2}$ is 
a chiral primary field, while $\varphi_{h=1,q=-2}$ is anti-chiral, we have
\be
G(z) \left[\varphi_{h=1,q=2} + \varphi_{h=1,q=-2}\right] = \frac{1}{z-x} \Bigl[
G^-_{-1/2} \varphi_{h=1,q=2} +
G^+_{-1/2} \varphi_{h=1,q=-2} \Bigr] \ + \ \hbox{regular} \ ,
\ee
which therefore cancels precisely (\ref{bulkglueshift}). This ties in
precisely with 
what we found in the field theory approach where the boundary term
(\ref{Bdef1}) restored the $\N=1$ supersymmetry.

To leading order this perturbation respects the conformal symmetry (since
the boundary field is marginal). However, generically, the boundary field in 
(\ref{Sbd}) is not exactly marginal, and thus this perturbation will induce an
RG flow to higher order. This should correspond to the RG flow of
\cite{Baumgartl:2007an}.

\subsection{$\N=2$ B-type branes under (ac) deformations}

It is expected \cite{Brunner:1999jq} that the stability of B-type branes depends
on the (ac) moduli. This has been made more precise in 
\cite{Douglas:2000gi,Aspinwall:2001dz,Walcher:2004tx} where
it was shown that (ac) perturbations can change the boundary condition
for the spectral flow operator and hence possibly destroy the charge
quantisation condition in the open string sector. As a consequence, a
(ac) bulk perturbation may lead to the appearance of open string tachyons, 
triggering a decay of the brane at lines of marginal stability. Note that
the perturbation by (ac) fields is rather different in nature than that by
(cc) fields: the former influence the integer charge quantisation and
lead to a potential breaking of conformal invariance, while world-sheet 
supersymmetry is always preserved. The latter
can potentially break world-sheet supersymmetry, leading to obstructions,
but leave the integer charge quantisation intact. 
\smallskip

In the following we want to explain how 
these effects appear from our point of view.
As before, the first step consists of analysing whether
the bulk perturbation by $\widehat\Phi$ induces a marginal (or
relevant) boundary field as it approaches the boundary. (If this
happens, then the bulk perturbation will break the conformal
invariance and hence induce an RG flow.)
This question is now controlled by the fusion of a chiral primary field
$\phi_c$, with an anti-chiral primary field $\bar{\phi}_a$. Using the fusion
rules of the $\N=2$ algebra \cite{Mussardo:1988av,Gaberdiel:1993mt},
one can show that only the even fusion rule can contribute, 
\be\label{fuac}
\phi_c \otimes \bar{\phi}_a = [\varphi_{q=0} ] \ ,
\ee
but one cannot deduce any constraints about the conformal dimension
of $\varphi$ in general.\footnote{We shall see in the examples below that
non-trivial relevant fields do indeed appear.} 
We therefore cannot exclude that such a flow
is induced by the bulk perturbation. In fact, this flow is the world-sheet 
description of what happens on a line of marginal stability, and the line of 
marginal stability is the location in moduli space where the 
boundary field that is switched on by the bulk perturbation becomes
marginal. As one crosses the line of marginal stability, the brane
undergoes a non-trivial RG flow, and thus decays. At the end-point 
of the RG flow the would-be marginal boundary field is no longer
marginal, and the new configuration is stable against further (ac)
deformations. 
\smallskip

Let us thus assume that we consider a configuration in the interior
of the moduli space, away from any line of marginal stability. This means
that (apart from the vacuum whose contribution we can regularise) 
the fusion rules (\ref{fuac}) do not contain any relevant or marginal fields.
It is then clear that (again up to possible vacuum contributions) the
field $\widehat\Phi$ vanishes in the limit as it approaches the boundary. 

We can then study the behaviour of the gluing conditions under the 
perturbation. The analysis is very similar to what was done before; in 
particular we now find that the potential contribution would have to come 
from $G^-_{-1/2} \varphi_{q=0}$ (for $\Delta G^+$) and 
$G^+_{-1/2} \varphi_{q=0}$ (for $\Delta G^-$). However, by our assumption
above, neither of these fields survives in the $y\rightarrow 0$ limit. Thus we 
conclude that none of the gluing conditions are modified, so that
no obstruction occurs.  This explains, from a world-sheet point of view,
why (ac) perturbations of B-type branes do not break supersymmetry, 
in agreement with the above expectations.

\subsection{The $\N=4$ analysis}

After this $\N=2$ discussion we now want to understand to which extent 
these statements are modified 
for $\N=4$ superconformal branes. Before we discuss the details, there is 
one general point we should stress. Suppose we are given some boundary
condition that preserves the $\N=4$ superconformal algebra up to some
automorphism. We may absorb this automorphism into a redefinition of the 
left- or right-moving $\N=4$ generators, and hence assume that the boundary 
condition preserves the $\N=4$ algebra without any automorphism. With respect
to the usual $\N=2$ subalgebra, the resulting D-brane is then a B-type D-brane. 

We are interested in studying the behaviour of this D-brane under bulk
deformations (that preserve the $\N=(4,4)$ superconformal symmetry in 
the bulk). As we have mentioned before, such bulk deformations are of four
different types: (cc), (ac,), (ca) or (aa). In an $\N=(2,2)$ theory these four
fields lie in {\em different} $\N=(2,2)$ multiplets; however in the $\N=(4,4)$ case
they always combine into one common $\N=(4,4)$ multiplet. 

One may now be tempted to believe that the $\N=(4,4)$ symmetry would
guarantee that the above brane behaves the same way under a (cc)
perturbation, as under an (ac) perturbation, say. However, this is not
true. Once we consider a brane that preserves the $\N=2$ subalgebra
without any automorphism (and hence is a B-type brane with respect to
the $\N=2$ subalgebra) then the $\N=4$ automorphisms that preserve
this property are the `outer automorphisms' of the $\N=4$ algebra that
rotate the $G^\pm$ and $G^{'\pm}$ modes into one another. However,
these automorphisms do {\em not} map a (cc) field to an (ac) field
(since they do not modify the separate U(1)-charges). Put differently,
the $\N=(4,4)$ automorphism that relates a (cc) perturbation to an
(ac) perturbation also maps a B-type brane (with respect to the $\N=2$
subalgebra) to an A-type brane (with respect to the same $\N=2$
algebra). Thus, even in the situation with $\N=(4,4)$ supersymmetry,
there are two cases to consider. We shall now discuss them in turn.

\subsubsection{(cc) perturbation of B-type branes}

To start with we just employ the $\N=(2,2)$ superconformal
symmetry. Thus we can repeat the analysis of section~3.2. In
particular, it therefore follows that the (cc) perturbation will not
break the conformal symmetry. On the other hand, as we saw in
section~3.2.1, generically the $\N=2$ superconformal symmetry may be
broken. In fact, the term that modifies the gluing condition for the
$G^+$ generator is precisely given by (\ref{Gpch}). Note that
this term only breaks the symmetry if the right-hand-side of
(\ref{Gpch}) is {\em not} part of the symmetry algebra of the theory,
{\it i.e.}\ if $\varphi_{h=1,q=2}$ is not part of the symmetry
algebra. Otherwise, (\ref{Gpch}) only means that the gluing condition
for $G^+$ is suitably modified.

If we are just considering an $\N=2$ superconformal field theory, then
the chiral algebra does not contain a field of $h=1$ and $q=2$, and
hence the symmetry is broken (provided that $\varphi_{h=1,q=2}\neq
0$). In fact, {\em if} \ the $\N=2$ superconformal field theory
contains such a chiral field (as well as its hermitian conjugate),
then the symmetry is enlarged, and one finds that the resulting chiral
algebra must contain an algebra with $\N=4$ superconformal symmetry
(where the additional supercharges are obtained from $G^\mp$ by the action of 
$\varphi_{h=1,q=\pm 2}\cong J^{\pm}$). Conversely, if the boundary theory 
preserves the $\N=4$
superconformal symmetry for $c=6$, then the representation theory of
the $\N=4$ algebra at $c=6$ $(k=1)$ implies that $\varphi_{h=1,q=2}
\cong J^+$ \cite{Eguchi:1988af}. Thus in this case, the gluing
condition (\ref{Gpch}) is modified by a field in the chiral algebra
itself.\footnote{For $\varphi_{h=1,q=2} = J^+$, the field 
$G^{-}_{-1/2}J^+$ is another supercurrent (namely 
$G^{-}_{-1/2}J^+ = - {G'}^{+}$), as follows from the $\N=4$ commutation relations
(\ref{N4re}).}  Then the symmetry is not broken but only the gluing condition
is modified. Since the same argument also holds for ${G'}^+$ (and
similarly for $G^-$ and ${G'}^-$), it then follows that the full
$\N=4$ superconformal symmetry is preserved in this case.  This shows
that $\N=4$ B-type branes are never obstructed against (cc)
perturbations. This is in nice agreement with what we found in the
superspace analysis of section~2.

Although the $\N=4$ superconformal symmetry is preserved, the gluing
condition for the $\N=1$ subalgebra is generically broken. This can be
adjusted as in section~3.2.2.  In the $\N=4$ case, the boundary field
in (\ref{Sbd}) is exactly marginal, and hence no higher order RG flow
is induced.

\subsubsection{(ac) perturbations of B-type branes}

The situation where we perturb an $\N=4$ B-type brane by an (ac)
perturbation is more subtle. From the point of view of the $\N=2$
analysis, we saw that the bulk perturbation may switch on a marginal
field on the boundary. The geometric interpretation of this phenomenon
was that this happens on a line of marginal stability. One may be
tempted to believe that the $\N=4$ representation theory should imply
that such a phenomenon cannot happen. However, this does not seem to
be the case (see appendix~B.2). In fact, there is an explicit example
(see section~4.2) where a non-trivial boundary field on an $\N=4$
preserving D-brane is switched on by an (ac) bulk deformation. At
least in that example, this boundary field will induce an RG flow on
the boundary (and thus modify the D-brane significantly) but it does
not seem to lead to a `brane decay' and therefore should probably not
be interpreted as indicating a `line of marginal stability'. However,
from the point of view of conformal perturbation theory, the fact that
no `brane decay' takes place cannot be seen at leading order in
perturbation theory, and it is therefore difficult --- even in the
situation with $\N=4$ --- to make definite general predictions about
the absence or otherwise of possible brane decays for $\N=4$ branes.

Finally, if we sit at a generic point in the (ac) moduli space, {\it
  i.e.}\ not on a line of marginal stability, then the same conclusion
as in section~3.3 applies, and we conclude that there is no
obstruction.

\section{Some examples}\label{Sect:K3example}
\setcounter{equation}{0}

In this section we want to present two simple examples that illustrate
some of the claims we have been making above. In particular, we want
to exemplify the mechanisms behind the obstruction of B-type $\N=2$
branes under (cc) deformations in a simple torus example
(section~4.1). In section~4.2 we shall then discuss $\N=4$ branes on
K3 at the Gepner point and show that under (ac) perturbations
non-trivial boundary fields may be switched on.

\subsection{A simple torus example}

To illustrate some aspects of our $\N=2$ analysis, let us consider the
free theory on $T^4=T^2\times T^2$, where initially both tori are
taken to be square tori of the same size.  On each torus there is a
complex boson and a complex fermion whose (left-moving) modes we
denote by $\alpha^{(i)}_n$, $\cc\alpha^{(i)}_n$ and $\psi^{(i)}_m$,
$\cc \psi^{(i)}_m $, where $i=1,2$ labels the two tori. The
corresponding right-moving modes will be denoted by a tilde. All of
these modes satisfy the usual (anti)-commutation relations. The theory
has actually $\N=4$ supersymmetry --- see \eg \cite{Brunner:2006tc}
for the construction of the corresponding superconformal
algebra. Although this means that there will be no actual
obstructions, all the effects described above occur.  
\smallskip

\noindent We are interested in the  gluing condition determined by 
\be
\begin{array}{rclrcl}
\tilde{\alpha}^{(1)} & = &   \omega_\lambda(\alo) \ , \qquad &
\tilde{\bar{\alpha}}^{(1)} & = &   \omega_\lambda(\bar\alpha^{(1)}) \\
\tilde{\alpha}^{(2)} & = &   \omega_\lambda(\alpha^{(2)}) \ , \qquad &
\tilde{\bar{\alpha}}^{(2)} & = &   \omega_\lambda(\bar\alpha^{(2)}) \ ,
\end{array} 
\ee
where $\omega_\lambda$ is defined as 
\be \label{gluingGeneral}
\begin{array}{cc}
\omega_\lambda(\alo) = \cos 2\theta_\lambda \calo + \sin
2\theta_\lambda \alt & 
\omega_\lambda(\calo) = \cos 2\theta_\lambda \alo 
+\sin 2\theta_\lambda \calt \\
\omega_\lambda(\alt) = \cos 2\theta_\lambda \calt -\sin 2\theta_\lambda
\alo \
& \omega_\lambda(\calt) = \cos 2\thla \alt -\sin 2\theta_\lambda \calo \ .
\end{array}
\ee
We also impose similar gluing conditions for the fermions. For 
the initial square tori $\theta_\lambda=\tfrac{\pi}{4}$ is an allowed boundary 
condition, and one easily shows that it satisfies B-type gluing conditions 
for the usual (diagonal) $\N=2$ algebra
(see {\it e.g.}\ \cite{Brunner:2006tc}). 
\smallskip

\noindent Let us now perturb the theory by the (cc) and (aa) fields
\be
\bPhi_{cc} = \cpsio_{-1/2} \tilde{\bar{\psi}}^{(1)}_{-1/2}|0\rangle\ , \qquad
\bPhi_{aa} = \psio_{-1/2} \tilde{\psi}^{(1)}_{-1/2}|0\rangle\ ,
\ee
so that the full perturbation fields are
\be\label{pbul}
\begin{array}{rcl}
\GPhi_{cc} & = & G^-_{-1/2} \tilde{G}^-_{-1/2} \bPhi_{c,c} =
  -2\calo_{-1} \tilde{\bar{\alpha}}^{(1)}_{-1} |0\rangle \ , \\
\GPhi_{aa} & = & G^+_{-1/2}\rmv G^+_{-1/2} \bPhi_{a,a} =
  -2\alo_{-1}\tilde{\alpha}^{(1)}_{-1} |0\rangle \ .
\end{array}
\ee
The corresponding hermitian fields are then described by the two
linear combinations 
\begin{align}
&\delta S_1=\int d^2z\, (\GPhi_{cc}+\GPhi_{aa})\ , 
&&\text{and} &&
\delta S_2=i\int d^2z\, (\GPhi_{cc}-\GPhi_{aa})\ .
\end{align}
Physically, in terms of the moduli of the torus, $\delta S_1$
corresponds to switching on the combination $(g_{11}-g_{22})$ of the
diagonal elements of the first torus metric, while $\delta S_2$ corresponds to
modifying $g_{12}$ of the first torus.\footnote{The remaining moduli of the first torus 
(the volume form and the anti-symmetric B-field) correspond to deformations 
involving the (ca) and (ac) fields. In addition there are also similar moduli
corresponding to changing the second torus.} 

Given the explicit form of the perturbing fields (\ref{pbul}) as well
as the boundary  conditions (\ref{gluingGeneral}), it is easy to see
that $\Phi_{cc}$ only switches on  a boundary field with $h=2$ as it
approaches the boundary, and likewise for  
$\Phi_{aa}$. Thus the perturbation does not break the conformal invariance,
in agreement with the analysis of section~3.2. The effect on the
gluing map is  very similar to what has been analysed in
\cite{Fredenhagen:2007rx}. To first order  it equals
\be
\delta \om(\alo) = 2 \cos 2\thla \alt \ ,
\ee
and actually this can be integrated up for arbitrary finite
$\lambda$. This leads to  the differential equation
\be
\frac{d}{d\lambda}\om(\alo) = \pi \sin^2 2\thla \calo - \pi\cos 2\thla
\sin 2\thla \alt 
\ ,
\ee
which is solved by $\omega_\lambda(\alo)$ if $\thla$ satisfies 
\be\label{diffeqtheta}
\dot \thla = -\frac{\pi}{2}\sin 2\thla\ .
\ee
Similar differential equations can be obtained for the other bosonic
modes as well, and all of them just correspond to modifying $\omega_\lambda$ 
as in (\ref{diffeqtheta}).

It is also straightforward to see that the gluing condition for
the fermions is not affected by the perturbation. Since the $\N=2$ 
supercharges are bilinear combinations of bosons and fermions, it
is then straightforward to determine their gluing conditions. One finds
that for $\lambda\neq 0$ they do not close any longer on the $\N=2$
subalgebra, but rather involve also the other supercharges of the 
$\N=4$ algebra \cite{Brunner:2006tc}. Among other things this then
also implies that the $\N=1$ gluing conditions are modified. To restore them
we proceed as in section~3.2.2. and switch on the boundary perturbation
(\ref{Sbd}), which in the current context takes the form
\be
i\lambda \int_\mathbb{R} dx\, \left(\bPhi_{c,c}(x)-\bPhi_{a,a}(x)\right) = 
i\lambda \int_\mathbb{R} dx\, \left( \psio \om(\psio) + \cpsio \om(\cpsio) 
\right)  \ .
\ee
This field induces a change in the gluing condition for the fermions 
(while it does not affect the bosons); following \cite{Recknagel:1998ih}, 
it leads to
\be
\delta\om(\psi) = -2\pi \delta\lambda \left[\Bigl(\psio \om(\psio) + \cpsio
\om(\cpsio)\Bigr)_0, \om(\psi) \right] \ .
\ee
In particular, this therefore guarantees that the fermions continue to satisfy
the same gluing condition as the bosons, {\it i.e.}\ the one determined in 
terms of $\omega_\lambda$. It is then in particular manifest that the original
$\N=1$ is preserved. While this restores the $\N=1$ gluing conditions, it does
not do so for the $\N=2$ B-type gluing conditions. However, the resulting
boundary conditions still preserves the $\N=4$ symmetry 
\cite{Brunner:2006tc}, in agreement with the analysis of section~3.4.1.

\subsection{K3 at the Fermat point} \label{Gepner}
\setcounter{equation}{0}

Another interesting example is the non-linear sigma model on K3, whose
infrared theory defines an $\N=(4,4)$ superconformal theory. More precisely,
we will concentrate on the Fermat point given by the zero locus of 
$W=x_1^4+x_2^4+x_3^4+x_4^4$, where there are
(at least) two alternative formulations: as a Gepner model 
\cite{Gepner:1987qi, Gepner:1989gr}, associated to
four copies of the $k=2$ $\N=2$ minimal model, or as the torus orbifold
$T^4/{\mathbb Z}_4$. In the Gepner construction only 
the $\N=(2,2)$ superconformal symmetry is manifest; the additional 
$SU(2)$ currents that enhance it to $\N=(4,4)$ are explicitly given as 
\begin{align}
&J^+ = \bigotimes_{i=1}^r (0,2,2)\otimes \overline{(0,0,0)}=
\bigotimes_{i=1}^r (k_i,-k_i,0)\otimes \overline{(0,0,0)}\ ,\label{defJ^+}\\
&J^-=\bigotimes_{i=1}^r (0,-2,2)\otimes \overline{(0,0,0)} =
\bigotimes_{i=1}^r (k_i,k_i,0)\otimes \overline{(0,0,0)} \ ,\label{defJ^-}
\end{align}
and similarly for the right-movers. Here, as in the following, we shall use
the conventions of \cite{Brunner:2006tc}.

We shall consider the B-type D-brane that 
is formulated as a permutation brane \cite{Recknagel:2002qq} associated to the 
$(12)(34)$ permutation. In the orbifold description, this brane corresponds
to a superposition of two D2-branes, one with Neumann directions
along $x^1$ and $x^3$, and the other with Neumann directions along 
$x^2$ and $x^4$. This configuration is then invariant under the 
${\mathbb Z}_4$ rotation action. (For details about these branes and their
identification see \cite{Brunner:2006tc}.) Since this boundary condition actually 
preserves the $\N=4$ superconformal symmetry, we should be able to test our 
predictions of section~3.4. 
\smallskip

First of all, we have checked that the $h=1, q=2$ field that is potentially
switched on by the (cc) perturbations is always the $J^+$ current on 
the boundary, {\it i.e.}\ the restriction of the bulk $J^+$ current (\ref{defJ^+}) to the 
boundary. This shows that this brane is not obstructed against any of the (cc)
perturbations. 

Incidentally, a similar analysis for an $\N=2$ brane
in a $c=9$ Calabi-Yau Gepner model typically leads to a boundary
field that is not exactly marginal. For example, consider 
the $(12)(35)(4)$ brane in the  $(k=3)^{\otimes 5}$ model, and 
perturb by the (cc) field  
\ba
\bPhi &=& 
\Bigl( (3,-3,0)\otimes \overline{(3,-3,0)} \Bigr) \otimes
\Bigl( (0,0,0)\otimes  \overline{(0,0,0)} \Bigr) \otimes
\Bigl( (2,-2,0)\otimes \overline{(2,-2,0)}\Bigr) \nonumber \\
& & \qquad  \qquad \otimes
\Bigl( (0,0,0) \otimes \overline{(0,0,0)} \Bigr) \otimes
\Bigl( (0,0,0) \otimes \overline{(0,0,0)}\Bigr) \ .
\ea
On the permutation brane this bulk field induces the boundary field
\be
\Psi = (3,-3,0)\otimes (3,-3,0)\otimes (2,-2,0)\otimes (0,0,0) \otimes
(2,-2,0)\ ,
\ee
which of course has $h=1$ and $q=2$. The OPE of $\Psi$ with itself
then contains
\be
\Psi\Psi \sim (3,-1,2) \otimes (3,-1,2) \otimes (3,1,2) \otimes (0,0,0)
\otimes (3,1,2) \ ,
\ee
which has $h=\frac{14}{5}$, thus showing that $\Psi$ does not belong
to the chiral algebra, {\it i.e.}\ that it is not exactly marginal. We would therefore
expect that a non-trivial boundary flow is induced, and this is indeed what was
found in \cite{Baumgartl:2007an}.
\medskip

Let us now return to the case of the $(12)(34)$ permutation brane on the 
$(k=2)^{\otimes 4}$ model of K3, and analyse the behaviour under an
(ac) perturbations. As an example of such a perturbation we consider the bulk 
field
\be
\phi = \Bigl( (0,0,0)\otimes \overline{(0,2,2)} \Bigr) \otimes
\Bigl( (1,1,0) \otimes \overline{(1,3,2)} \Bigr) \otimes
\Bigl( (1,1,0) \otimes \overline{(1,3,2)} \Bigr) \otimes
\Bigl( (2,2,0) \otimes \overline{(2,4,2)} \Bigr) \
\ee
from the first twisted sector of the Gepner orbifold. (From the point of view
of the torus orbifold, this field arises in the first twisted sector
of the ${\mathbb Z}_4$ orbifold.) Again, using the permutation gluing condition,
it is easy to see that $\phi$ induces the boundary field 
\be
\Psi = (1,-1,0) \otimes (1,-1,0) \otimes (1,1,0) \otimes (1,1,0)
\ee
on the boundary. This field has $q=0$ and $h=1/2$, and thus is an example of 
a non-trivial (relevant) field appearing in the fusion rule
(\ref{B11}). From the orbifold point of view, this field has also a
simple interpretation: it is precisely the lowest mode of the open
string between the two D2-branes. (Note that from the point of  view
of the orbifold theory this is not a boundary changing field since
only the  superpositions of D2-branes is orbifold invariant. In fact,
since the bulk field comes from 
the first twisted sector, it transforms with a primitive fourth root
of unity under the  
orbifold quantum symmetry. The corresponding boundary field must then transform 
with the complex conjugate phase under this quantum symmetry since otherwise
the bulk-boundary correlator would vanish. But this means that it must
be a boundary changing field since these get projected out when one performs
the quantum symmetry orbifold to return to the original theory.)

Geometrically, this boundary field will probably modify the way the
two D2-branes meet at the orbifold singularity --- the bulk field in
question describes after all one of the blowing-up modes of the
orbifold fixed point. However, we do not think this will lead to any
form of `brane decay' since there are probably no configurations of
two D-branes that have the same RR charge and tension. Thus we should
probably not interpret this as an example of a line of marginal
stability. However, from the perturbed conformal field theory point of
view, this phenomenon looks identical to what happens, say, in the
$\N=2$ context where there are lines of marginal stability.

\section{Conclusions}

In this paper we have studied how the obstructions of supersymmetric
branes may be understood from a world-sheet perspective. In particular,
we have explained how the (possible) obstruction of an $\N=2$ B-type brane
under a (cc) perturbation arises in the superspace field theory language, and 
from the point of view of perturbed conformal field theory.
We have also explained that this
phenomenon does not occur for $\N=4$ superconformal branes at $c=6$, 
in particular for half-supersymmetric D-branes on K3.

While our analysis for the perturbations of B-type branes under (cc)
deformations is fairly complete, the behaviour of B-type branes under
(ac) deformations (or A-type branes under (cc) deformations) is more
subtle. In the $\N=2$ context it is believed that there are no actual
obstructions, but that branes may decay along lines of marginal
stability. This phenomenon is not visible in the superspace field
theory language. However, we can understand the origin of it in the
conformal field theory analysis where it corresponds to the situation
when an irrelevant boundary field (that is switched on by the bulk
deformation) becomes marginal (see section~3.3). In the
case of half-BPS branes in $\N=4$ theories, there is evidence
that lines of marginal stability do not exist. This expectation is
mainly based on the fact that the index that counts half-BPS states
is constant on the full moduli space.
However, the analysis of section~4.2 seems to suggest that
at least relevant boundary fields may be switched on by bulk
deformations, even in the $\N=4$ case. Thus on the basis of our
leading order conformal field theory analysis it does not appear to be
possible to detect any difference between the $\N=2$ and $\N=4$ case
in this respect. It would be interesting to understand whether the
higher order behaviour is different between the $\N=2$ and the $\N=4$
case.  It would also be very interesting to relate this world-sheet
analysis with recent progress for `wall-crossing formulae' of $\N=2$
branes, see for example \cite{Gaiotto:2008cd}.

\section*{Acknowledgements} 
We thank Costas Bachas for collaboration at an early stage of this
project. We also thank Miranda Cheng,  Sergei Gukov, Kentaro Hori, 
Andy Neitzke and Suresh Nampuri for 
helpful discussions, and Ashoke Sen for a useful communication. 
The research of I.B. is supported by a EURYI award, while the research
of M.R.G. and S.H.  
is supported by the Swiss National Science Foundation. ETH and LMU were
nodes of the Marie Curie network `Constituents, Fundamental Forces and Symmetries
of the Universe' (MRTN-CT-2004-005104). C.A.K. is supported by a Fellowship of 
the Swiss National Science Foundation.

\appendix

\section{Superspace conventions}
\renewcommand{\theequation}{\Alph{section}.\arabic{equation}}
\setcounter{equation}{0}
\subsection{$\N=2$ superspace conventions}\label{App:SuperspaceN2}
The $\N=2$ supersymmetry transformations are generated by the
following supercharges 
\begin{align}
\Q_\pm =\frac{\partial}{ \partial \theta^\pm} + i \bar\theta^\pm
\partial_\pm\ , 
&& &\bar\Q_\pm  = - \frac{\partial}{ \partial \bar\theta^\pm} 
- i \theta^\pm  \partial_\pm \ . \nonumber
\end{align}
Here $\partial_\pm = \tfrac{1}{2} (\partial_0 \pm \partial_1)$ and 
$x^\pm = x^0 \pm x^1$ are the usual light-cone coordinates. One easily
checks that the only non-trivial anti-commutators of these generators
are 
\be
\{ \Q_\pm  , \bar\Q_\pm  \} = - 2 i \partial_\pm \ . 
\ee
For the following it is also useful to introduce the corresponding
spinor derivatives 
\begin{align}
&\D_\pm=\frac{\partial}{ \partial \theta^\pm} - i \bar\theta^\pm \partial_\pm\ , 
&& \bar\D_\pm = - \frac{\partial}{ \partial \bar\theta^\pm} 
+ i \theta^\pm  \partial_\pm \ . \nonumber
\end{align}
One easily confirms that all of these generators anti-commute with all
the $\Q_\pm $ and $\bar\Q_\pm $. Furthermore, the only non-trivial
anti-commutator involving the $\D$ generators is 
\be
\{ \D_\pm , \bar\D_\pm \} = + 2 i \partial_\pm \ . 
\ee
With the help of these operators we can now define what we mean by a
chiral superfield $\Phi$: this is a superfield that satisfies
\be\label{chdef}
\bar\D_\pm \Phi = 0 \ .
\ee
Every chiral superfield can be written as 
\be\label{chsup}
\Phi(x^\pm,\theta^\pm,\bar\theta^\pm) = 
\phi(y^\pm) + \theta^+ \psi_+ (y^\pm) + \theta^- \psi_- (y^\pm)+ 
\theta^+ \theta^- F(y^\pm) \ , 
\ee
where 
\be\label{ydef}
y^\pm = x^\pm - i \theta^\pm \bar\theta^\pm \ . 
\ee
Indeed, one easily checks that $\Phi$ in (\ref{chsup}) satisfies 
(\ref{chdef}). The complex conjugate of a chiral superfield $\Phi$ is
an anti-chiral superfield $\bar\Phi$; anti-chiral superfields are
characterised by $\D_\pm \bar\Phi=0$, and they have an expansion
\be\label{achsup}
\bar\Phi(x^\pm,\theta^\pm,\bar\theta^\pm) = 
\bar\phi(\bar{y}^\pm) + \bar\theta^+ 
\bar\psi_+(\bar{y}^\pm) + \bar\theta^- 
\bar\psi_-(\bar{y}^\pm)+  
\bar\theta^+ \bar\theta^- \bar{F}(\bar{y}^\pm) \ , 
\ee
where 
\be
\bar{y}^\pm = x^\pm + i \theta^\pm \bar\theta^\pm \ . 
\ee
A twisted chiral superfield $U$ is a superfield that satisfies 
\be
\bar\D_+ U=\D_- U = 0 \ . \label{N2twist}
\ee
Every twisted chiral superfield can be written as
\be
U(x^\pm,\theta^{\pm} ,\bar\theta^{\pm} ) = 
v(\tilde{y}^\pm) +\theta^+ \chi_+(\tilde{y}^\pm) 
+ \bar\theta^- \bar\chi_-(\tilde{y}^\pm)
+\theta^+\bar\theta^- E(\tilde{y}^\pm)\ ,
\ee
where
\be
\tilde{y}^\pm = x^\pm \mp i \theta^\pm \bar\theta^\pm \ .
\ee
Finally, a general twisted anti-chiral superfield that is
characterised by 
\be
\D_+ \bar{U}=\bar\D_- \bar{U} = 0 \label{N2antitwist}
\ee
can always be written as 
\be
\bar{U}(x^\pm,\theta^{\pm} ,\bar\theta^{\pm} ) = 
\bar{v}(\hat{y}^\pm) +\bar\theta^+ \bar\chi_+(\hat{y}^\pm) 
+ \theta^- \chi_-(\hat{y}^\pm)
+\bar\theta^+ \theta^- \bar{E}(\hat{y}^\pm)\ ,
\ee
where
\be
\hat{y}^\pm = x^\pm \pm i \theta^\pm \bar\theta^\pm \ .
\ee

\subsection{$\N=4$ superspace conventions}\label{App:Superspace4}
In what follows it will be understood that shifting of $SU(2)$ indices is
accomplished 
with the help of $\epsilon$-symbols. 
In the $\N=4$ standard superspace (\ref{standardss}), supersymmetry 
transformations will be generated by
\begin{align}
&\Q^i=\frac{\partial}{\partial\theta_i}+i\bth^i\partial_+\ , 
&&\text{and} &&\bar{\Q}_i
=-\frac{\partial}{\partial\bth^i}-i\theta_i\partial_+\ ,\\
&\Q^{\bar{\imath}}=\frac{\partial}{\partial\theta_{\bar{\imath}}}
+i\bth^{\,\bar{\imath}}\partial_-\ , 
&&\text{and} &&\bar{\Q}_{\bar{\imath}}=
-\frac{\partial}{\partial\bth^{\bar{\imath}}}
-i\theta_{\bar{\imath}}\partial_-\ ,\label{N4standardSC}
\end{align}
with the following non-trivial anti-commutation relations
\begin{align}
&\{\Q^i,\bar{\Q}_j\}=-2i\delta^i_j\partial_+\ , 
&&\{\Q^{\bar{\imath}},\bar{\Q}_{\bar{\jmath}}\}=
-2i\delta^{\bar{\imath}}_{\bar{\jmath}}\partial_- \ .\label{SC4commrel}
\end{align}
Similarly, we can also introduce the corresponding spinor derivatives
which take the form 
\begin{align}
&\D^i=\frac{\partial}{\partial\theta_i}-i\bth^i\partial_+\ , 
&&\text{and} &&\bar{\D}_i=
-\frac{\partial}{\partial\bth^i}+i\theta_i\partial_+\ ,\\
&\D^{\bar{\imath}}=\frac{\partial}{\partial\theta_{\bar{\imath}}}
-i\bth^{\bar{\imath}}\partial_-\ , 
&&\text{and} &&\bar{\D}_{\bar{\imath}}
=-\frac{\partial}{\partial\bth^{\bar{\imath}}}
+i\theta_{\bar{\imath}}\partial_-\ .\label{N4spinorder}
\end{align}
The only non-vanishing anti-commutation relations involving $\D$s are
given by 
\begin{align}
&\{\D^i,\bar{\D}_j\}=2i\delta^i_j\partial_+\ , 
&&\{\D^{\bar{\imath}},\bar{\D}_{\bar{\jmath}}\}=
2i\delta^{\bar{\imath}}_{\bar{\jmath}}\, 
\partial_-\ .\label{SD4commrel}
\end{align}
In order to avoid cluttering our formulae we combine all spinors into
doublets with respect to $SU(2)_{c}$. First of all, we write the
Grassmann variables as\footnote{Notice that they transform as
$(\mathbf{2},\mathbf{2})$ under $SU(2)_{c}\times SU(2)_{f}$.}
\begin{align}
&\theta_i^{a}=\left(\begin{array}{c}\theta_i \\ \bar{\theta}_i
\end{array}\right)\, ,&&\text{and} &&\theta_{\bar{\imath}}^{\bar{a}}
=\left(\begin{array}{c}\theta_{\bar{\imath}} \\ 
\bar{\theta}_{\bar{\imath}}\end{array}\right)\, ,\label{DoubletGrassmann}
\end{align}
as well as the supercharges 
\begin{align}
&\Q^i_a=\left(\begin{array}{c} \Q^i \\ 
\bar{\Q}^i\end{array}\right)\ , 
&&\Q^{\bar{\imath}}_{\bar{a}}=
\left(\begin{array}{c} \Q^{\bar{\imath}} \\ 
\bar{\Q}^{\bar{\imath}}\end{array}\right)
\ ,\label{SU(2)supercharge}
\end{align}
and finally also the spinor derivatives
\begin{align}
&\D^i_a=\left(\begin{array}{c} \D^i \\ \bar{\D}^i\end{array}\right)\ , 
&&\D^{\bar{\imath}}_{\bar{a}}=\left(\begin{array}{c} \D^{\bar{\imath}} \\ 
\bar{\D}^{\bar{\imath}}\end{array}\right).
\end{align}
An example of a superfield living on the superspace (\ref{standardss}) 
is the $\N=(4,4)$ twisted multiplet (see 
\cite{Gates:1984nk,Ivanov:1994er,Ivanov:2004re}) whose Grassmann 
expansion reads
\begin{align}
\Phi^{i\bar{\imath}}=&\varphi^{i\bar{\imath}}+2\theta^{a,i}\psi^{\bar{\imath}}_a
+2\theta^{\bar{a},\bar{\imath}}\psi^{i}_{\bar{a}}
+2\theta^{a,i}\theta^{\bar{a},\bar{\imath}}F_{a\bar{a}}
+\text{derivatives}\ .\label{standardsuperfield}
\end{align}
The lowest component of this field is given by a quartet of
(pseudo-real) scalar fields,  
which satisfy the relation
\begin{align}
\overline{(\varphi^{i\bar{\imath}})}=\epsilon_{ij}
\epsilon_{\bar{\imath}\,\bar{\jmath}} \, \varphi^{j\bar{\jmath}}\ .
\end{align}

\subsection{Harmonic superspace conventions}\label{App:SuperspaceHS}
\subsubsection{Supercharges and spinor derivatives}
In (\ref{trueharmonics}) we have introduced harmonic coordinates
\begin{align}
&\{\udpme{i}\}\in \frac{SU(2)_L}{U(1)_L}\,, &&\text{and} 
&&\{\udpmz{\bar{\imath}}\}\in \frac{SU(2)_R}{U(1)_R}\ ,
\end{align}
where $SU(2)_L$ and $SU(2)_R$ will be identified with the left- and right-moving $SU(2)_f$,
respectively. Explicitly we can write the harmonic variables in the following matrix form
\begin{align}\label{A25}
&\left(\begin{array}{cc} \udme{1} & \udpe{1} 
\\ \udme{2} & \udpe{2} \end{array} \right) \in SU(2)_L\ , 
&&\text{and} &&\left(\begin{array}{cc} \udmz{1} & \udpz{1} 
\\ \udmz{2} & \udpz{2} \end{array} \right) \in SU(2)_R \ .
\end{align}
Acting with $SU(2)_L$ on the left, we see that both $\udpe{i}$ and
$\udme{i}$ are  doublets. The $\pm$ indicates the $U(1)_L$ charge, as
can be seen by acting with $U(1)_L \subset SU(2)_L$ on the
right. Similar statements obviously hold for the right  moving
expressions. We define raising and lowering of indices via complex 
conjugation, 
\begin{align}
&\begin{array}{l}\overline{\uupe{i}}=\udme{i} \\ 
\overline{\udpe{i}}=-\uume{i}\end{array}
&&\text{and} && \begin{array}{l}
\overline{\uupz{\bar{\imath}}}=\udmz{\bar{\imath}} \\ 
\overline{\udpz{\bar{\imath}}}=-\uumz{\bar{\imath}} \ .\end{array}
\end{align}
It then follows that
\begin{align}
&\uupe{i}\udme{i}=1 &&\text{and} && \uupz{\bar{i}}\udmz{\bar{i}}=1\ , 
&&\text{(unit determinant condition)} \ ,\label{unitdet}
\end{align}
and, using the convention $\epsilon_{12}=-1$,
\begin{align}
&\epsilon_{ij}=\udpe{i}\udme{j}-\udme{i}\udpe{j}\ ,
&&\text{and} &&\epsilon_{\bar{\imath}\, \bar{\jmath}}=
\udpz{\bar{\imath}}\udmz{\bar{\jmath}}-\udmz{\bar{\imath}}
\udpz{\bar{\jmath}}\ .
\label{epsilonident}
\end{align}
These relations entail that not all of the harmonic variables are
independent of each other.  This has in particular the effect that
na\"ive partial derivatives with respect to $u$ are not
allowed. Instead one has to use covariant harmonic derivatives, which
take the following form (see for example
\cite{Ivanov:1994er,Galperin:2001uw}) 
\begin{align}
&D^{(\pm2,0)}=\udpme{i}\frac{\partial}{\partial \udmpe{i}}\ ,
&&D^{(0,\pm2)}=\udpmz{i}\frac{\partial}{\partial \udmpz{i}}\ ,\\
&D^{(0,0)}_{(L)}=\udpme{i}\frac{\partial}{\partial \udpme{i}}\ , 
&&D^{(0,0)}_{(R)}=\udpmz{i}\frac{\partial}{\partial \udpmz{i}}\ .
\end{align}
One can easily convince oneself that these derivatives respect the
relation (\ref{unitdet}).\\ 

\noindent
The integration over the harmonic variables is simply an integration
on the sphere  which picks the singlet piece of the harmonic expansion
of the integrand. In particular  
we have 
\begin{align}
\int \due1=\int \duz1=1\ ,\label{harmintI}
\end{align}
while we get on the other hand (for $m\neq0\neq n$)
\begin{align}
&\int \due u^{(+,0)}_{(i_1}\ldots u^{(+,0)}_{i_n}u^{(-,0)}_{j_1}
\ldots u^{(-,0)}_{j_m)}=0\ ,
\label{harmintII}\\
&\int \duz u^{(0,+)}_{(\bar{\imath}_1}\ldots u^{(0,+)}_{\bar{\imath}_n}
u^{(0,-)}_{\bar{\jmath}_1} \ldots u^{(0,-)}_{\bar{\jmath}_m)}=0 \ .
\end{align}
The supersymmetries in the formulation of (\ref{harmss}) are 
generated  by the supercharges of standard $\N=4$ superspace 
({\it i.e.}\ (\ref{N4standardSC})) projected by the harmonic
coordinates 
\begin{align}
&\Q^{(\pm,0)}_a=\udpme{i} \Q^i_a=\mp\frac{\partial}{\partial\theta^{a(\mp,0)}}
+i\bar{\theta}^{(\pm,0)}_a\partial_+\ ,\nonumber\\
&\Q^{(0,\pm)}_{\bar{a}}=\udpmz{\bar{\imath}} \Q^{\bar{\imath}}_{\bar{a}}=
\mp\frac{\partial}{\partial
\theta^{\bar{a}(0,\mp)}}+i\bar{\theta}^{(0,\pm)}_{\bar{a}}\partial_-\ .
\end{align}
Using (\ref{SC4commrel}), the projected commutation relations become
\begin{align}
&\{\Q^{(\pm,0)}_a,\Q^{(\mp,0)}_b\}=\mp \, 2i\, \epsilon_{ab}\, \partial_+\ , 
&&\{\Q^{(0,\pm)}_{\bar{a}},\Q^{(0,\mp)}_{\bar{b}}\}=
\mp \, 2i\, \epsilon_{\bar{a}\bar{b}}\,\partial_-\ .
\end{align}
By projecting (\ref{N4spinorder}) we can in the same manner also
define the projected spinor derivatives 
\begin{align}
&\D^{(\pm,0)}_a=\udpme{i} \D^i_a=\mp\frac{\partial}{\partial\theta^{a(\mp,0)}}
-i\bar{\theta}^{(\pm,0)}_a\partial_+\ ,\nonumber\\
&\D^{(0,\pm)}_{\bar{a}}=\udpmz{\bar{\imath}} \D^{\bar{\imath}}_{\bar{a}}
=\mp\frac{\partial}{\partial \theta^{\bar{a}(0,\mp)}}
-i\bar{\theta}^{(0,\pm)}_{\bar{a}}\partial_-\ ,
\end{align}
which (in view of (\ref{SD4commrel})) satisfy
\begin{align}
&\{\D^{(\pm,0)}_a,\D^{(\mp,0)}_b\}=\pm\,  2i\, \epsilon_{ab}\, \partial_+\ , 
&&\{\D^{(0,\pm)}_{\bar{a}},\D^{(0,\mp)}_{\bar{b}}\}=
\pm \, 2i\, \epsilon_{\bar{a}\bar{b}}\, \partial_-\ .
\end{align}
For further convenience let us also introduce the so called analytic
basis, in which we redefine the space-time coordinates in the
following manner 
\begin{align}
&z^+=x^+-i\theta^{a(+,0)}\bar{\theta}^{(-,0)}_a\ , 
&&\text{and} &&z^-=x^--i\theta^{\bar{a}(0,+)}
\bar{\theta}^{(0,-)}_{\bar{a}}\ .\label{analyticbasis}
\end{align}
In this basis, the supercharges become
\begin{align}
&\Q^{(+,0)}_a=-\frac{\partial}{\partial\theta^{a(-,0)}}
+2i\bar{\theta}^{(+,0)}_a\partial_+\ , 
&&\Q^{(-,0)}_a=\frac{\partial}{\partial\theta^{a(+,0)}}\ ,\nonumber\\
&\Q^{(0,+)}_{\bar{a}}=-\frac{\partial}{\partial \theta^{\bar{a}(0,-)}}
+2i\bar{\theta}^{(0,+)}_{\bar{a}}\partial_-\ , 
&&\Q^{(0,-)}_{\bar{a}}=\frac{\partial}{\partial
  \theta^{\bar{a}(0,+)}}\ ,
\label{analyticchar4}
\end{align}
while the spinor derivatives read
\begin{align}
&\D^{(+,0)}_a=-\frac{\partial}{\partial\theta^{a(-,0)}}\ , 
&&\D^{(-,0)}_a=\frac{\partial}{\partial\theta^{a(+,0)}}
-2i\bar{\theta}^{(-,0)}_a\partial_+\ ,
\nonumber\\
&\D^{(0,+)}_{\bar{a}}=-\frac{\partial}{\partial\theta^{\bar{a}(0,-)}}\ , 
&&\D^{(0,-)}_{\bar{a}}=\frac{\partial}{\partial \theta^{\bar{a}(0,+)}}
-2i\bar{\theta}^{(0,-)}_{\bar{a}}\partial_-\ .\label{analyticdiff}
\end{align}
Notice in particular that the operators
$(\D^{(+,0)}_a,\D^{(0,+)}_{\bar{a}})$ are simple partial derivatives
with respect to the Grassmann variables. 
\subsubsection{Superfields}
The superfields we are going to consider on (\ref{harmss}) are of the
following  type\footnote{We are working in the analytic basis
(\ref{analyticbasis}) and it is understood that all component fields
are functions of $z^\pm$ as well as the harmonic coordinates.}
\cite{Ivanov:1994er}  
\begin{align}
&\Phi^{(+,+)}=\fpp+2\tlpi{a}\chp_a+2\trpi{\bar{a}}\psp_{\bar{a}}
-i\tlpi{a}\tlpid{a}\partial_+\fmp
\nonumber\\
&\hspace{0.5cm}-i\trpi{\bar{a}}\trpid{\bar{a}}\partial_-\fpm
+\tlpi{a}\trpi{\bar{a}}F_{a\bar{a}}-2i\tlpi{a} \trpi{\bar{a}} 
\trpid{\bar{a}}\partial_-\chm_a
\nonumber\\
&\hspace{0.5cm}-2i\trpi{\bar{a}}\tlpi{a}\tlpid{a}\partial_+
\psm_{\bar{a}}-\tlpi{a}\tlpid{a}\trpi{\bar{a}}\trpid{\bar{a}}
\partial_+\partial_-\fmm\ .\label{superfield}
\end{align}
Here we are using harmonically projected component fields, for example
for the scalars 
\begin{align}
&\varphi^{(\pm,\pm)}=\varphi^{i\bar{\imath}}\udpme{i} \udpmz{\bar{\imath}}\ , 
\end{align}
while the notation for the fermions shall be
\begin{align}
&\psi^{(\pm,0)}_{\bar{a}}=\psi^i_{\bar{a}}\udpme{i}\ ,&&\text{and} 
&&\psi^{(0,\pm)}_a=\psi^{\bar{\imath}}_{a}\udpmz{\bar{\imath}}\ .
\end{align}
Notice in particular that the superfield (\ref{superfield}) has
charges $(+1,+1)$ with respect to $U(1)_L\times U(1)_R$. Moreover,
observe that the superfield $\Phi^{(+,+)}$ has an ultrashort component
expansion, which is due to our formulation of the theory on harmonic
superspace. To be more precise, $\Phi^{(+,+)}$ is G(rassmann)-analytic
in the sense that it only depends on half of all the Grassmann
variables. Formulated as a constraint on $\Phi^{(+,+)}$ this statement
means
\begin{align}
\D^{(+,0)}_a\Phi^{(+,+)}=\D^{(0,+)}_{\bar{a}}\Phi^{(+,+)}=0\ ,\label{analyticity}
\end{align}
which becomes apparent upon recalling (\ref{analyticdiff}). In a sense
one can understand equation (\ref{analyticity}) as the generalised
$\N=4$ version of the chirality conditions of $\N=2$ supersymmetry
({\it i.e.}\ equations (\ref{chdef}), (\ref{N2twist}) and
(\ref{N2antitwist})). Besides that $\Phi^{(+,+)}$ is also
H(armonically)-analytic, which means that the harmonic dependence of
its components is not arbitrary but satisfies
\begin{align}
D^{(+2,0)}\Phi^{(+,+)}=D^{(0,+2)}\Phi^{(+,+)}=0\ .\label{Hanalyticity}
\end{align}


\section{The $\N =2$ and $\N =4$ superconformal theories}\label{app:N4}
\setcounter{equation}{0}
\subsection{The $\N =2$ and $\N =4$ superconformal algebras}
The $\N=2$ superconformal
algebra is generated by the modes $L_n$ of the stress-energy tensor,
the modes $J_n$ of a $U(1)$-current, as well as by the modes of the
two supercurrents $G^\pm_r$. The commutation relations are 
\begin{align*}
  \left[L_m, L_n \right] &=(m-n)L_{m+n} + \tfrac{c}{12}(m^3 -  m)
      \delta_{m,-n} \ , \\ 
  \left[L_m, J_n \right] &=-nJ_{m+n}\ ,\\
  \left[L_m, G_n^\pm \right] &=\left( \tfrac{1}{2}m-n \right)
  G_{m+n}^\pm \ , \\
  \left[J_m, J_n \right] &= \tfrac{c}{3} m\delta_{m,-n}\ ,\\
  \left[J_m, G_n^\pm \right] &=\pm G_{m+n}^\pm \ , \\
  \left\{ G_{m}^+, G_{n}^-\right\} &= 2 L_{m+n} + (m-n) J_{m+n} +
  \tfrac{c}{3}(m^2 - \tfrac{1}{4})\delta_{m, -n} \ , \\
  \left\{G_{m}^+, G_{n}^+ \right\} &=
   \left\{G_{m}^-, G_{n}^-\right\}=0\ .
\end{align*}
Here the superscript $\pm$ denotes the $U(1)$ charge of the generator. In 
the $\N=4$ superconformal algebra the $U(1)$ symmetry is enhanced to an 
affine $SU(2)$ symmetry; the corresponding modes are denoted by $J^\pm_n$ and
$J_n$, and satisfy the commutation relations
\begin{eqnarray}
{}[J_m,J^\pm_n] & = & \pm \, 2 \, J^\pm_{m+n} \nonumber \\
{}[J^+_m,J^-_n] & = & J_{m+n} + \frac{c}{6}m \, \delta_{m,-n} \nonumber \\
{}[J_m,J_n] & = & \frac{c}{3} \, m \, \delta_{m,-n} \,. \nonumber
\end{eqnarray}
In addition there are  two more supercurrents, whose modes we denote by 
${G'}^\pm_r$. The additional commutation relations are 
\cite{Ademollo:1976wv}
\begin{eqnarray}
\{G^\pm_m,{G'}^\pm_n\} & = & \mp 2 (m-n) J^\pm_{m+n} \qquad \qquad 
\{G^\pm_m,{G'}^\mp_n\} =  0 \nonumber \\
{}[L_m,{G'}^\pm_n] & = & \left(\frac{m}{2} - n \right) {G'}^\pm_{m+n} 
\qquad \qquad 
{}[J_m,{G'}^\pm_n]  =  \pm {G'}^\pm_{m+n} \nonumber \\
{}[J^\pm_m,{G}^\pm_n] & = & [J^\pm_m,{G'}^\pm_n] = 0 \nonumber \\
{}[J^\pm_m,{G}^\mp_n] & = & \pm {G'}^\pm_{m+n} \qquad \qquad\qquad 
\quad\, 
{}[J^\pm_m,{G'}^\mp_n]  =  \mp {G}^\pm_{m+n} \nonumber \\
\{{G'}^{+}_m,{G'}^-_n\} & = & 2 L_{m+n} + (m-n) J_{m+n} + 
\frac{c}{3} (m^2 - \tfrac{1}{4}) \delta_{m,-n} \ . \label{N4re}
\end{eqnarray}
The relation of these operators to the supercharges of the superspace
approach of  appendix \ref{App:Superspace4} is
\begin{align}
\Q^i=\left(\begin{array}{c}{G'}^+_{-1/2} \\
    {G}^+_{-1/2}\end{array}\right)\ , 
&&\text{and} &&\bar{\Q}^i=\left(\begin{array}{c}G^-_{-1/2} \\ 
-{G'}^-_{-1/2}\end{array}\right)\ .
\end{align}
It is then straightforward to reproduce the action of
$SU(2)_c$ as well as the relation (\ref{SC4commrel}) 
\begin{align}
\{\Q^i,\bar{\Q}_j\}=2\delta^i_j L_{-1},
\end{align}
upon the identification of $L_{-1}=-i\partial_+$ with the generator of
infinitesimal bosonic translations.  

\subsection{Fusion rules for $\N=2$}\label{App:FusionN2}

In this appendix we summarise the fusion rules of chiral 
and anti-chiral primary fields. In general, there are three different
types  of fusion rules for $\N=2$ theories: the even fusion rules and
the two odd fusion rules \cite{Mussardo:1988av}. These are
characterised by the property that either  
\be
\langle \varphi , \phi_1 \otimes \phi_2 \rangle \neq 0 \qquad \hbox{even}
\ee
or 
\be
\langle \varphi , \left(G^\pm_{-1/2} \phi_1\right) 
\otimes \phi_2 \rangle \neq 0 
\qquad \hbox{$\pm$ odd}\ . 
\ee
Obviously, if $\phi_1$ is a chiral primary field, only the even
and $-$odd fusion rule can be non-trivial. This is also true if $\phi_2$ is 
a chiral primary, as can be seen from symmetry, or by using (\ref{comultG})
below.

First we want to show  that in the even fusion rule only one field can appear,
namely the chiral primary with $q=q_1+q_2$. It is clear by charge conservation
that $\varphi$ must have $q=q_1+q_2$; thus it suffices to show that
$\varphi$ must  be chiral primary. This follows because
\begin{eqnarray}
(2h-q) \langle \varphi , \phi_1\otimes \phi_2 \rangle & = & 
\langle G^-_{1/2} G^+_{-1/2} \varphi, \phi_1 \otimes \phi_2 \rangle  \nonumber \\
& = & 
\langle G^+_{-1/2} \varphi,  \, \Delta(G^+_{-1/2})(\phi_1 \otimes
\phi_2) \rangle = 0 \ , 
\end{eqnarray}
where $\Delta$ denotes the comultiplication \cite{Gaberdiel:1993mt} which in 
this case takes on the simple form
\be \label{comultG}
\Delta(G^+_{-1/2}) = G^+_{-1/2} \otimes {\bf 1} + 
{\bf 1} \otimes G^+_{-1/2}  \ .
\ee
This then vanishes, because both $\phi_1$ and $\phi_2$ are chiral
primaries.  In the case of interest, $h_1=h_2=\tfrac{1}{2}$ and
$q_1=q_2=1$, so that we have 
\be\label{cc}
\phi_c \otimes \phi_c = [ \varphi_{h=1,q=2}]  \oplus [G^+_{-1/2}
\varphi^+_{q=1} ] \ , 
\ee
where $\varphi^+$ has $h^+>\tfrac{1}{2}$, and $[\psi ]$ is the $\N=2$
representation generated from the state $\psi$.

\subsection{Fusion rules for $\N=4$}

Finally we want to collect some facts about the fusion rules of 
$\N=4$ representations.\footnote{As far as we are aware, this
problem has not been addressed before in the literature.}
We shall only consider the NS sector. Furthermore, we shall
only discuss the case that is of interest in the current context, 
namely the fusion of two `massless' representations with $j=\tfrac{1}{2}$
at $k=1$. The massless representations are those that saturate the
BPS bound, which in this context means that they have $h=\tfrac{1}{2}$. 
The Virasoro highest weight states of the representation with $j=\tfrac{1}{2}$ 
therefore forms a doublet $(\phi_c,\phi_a)$ with
\be
\begin{array}{rclrcl}
L_0 \phi_c & = & \frac{1}{2} \phi_c \qquad  
& L_0 \phi_a & = & \frac{1}{2} \phi_a \\ 
J_0 \phi_c & = & \phi_c \qquad 
& J_0 \phi_a & = & - \phi_a \\
J^+_0 \phi_c & = & 0 \qquad 
& J^+_0 \phi_a & = & \phi_c \\
J^-_0 \phi_c & = & \phi_a \qquad 
& J^-_0 \phi_a & = & 0 \ . 
\end{array}
\ee
With respect to the usual $\N=2$ subalgebra, $\phi_c$ is thus a chiral
primary state, while $\phi_a$ is an anti-chiral primary state. Note that 
the same is also true for the $\N=2$ subalgebra generated by 
${G'}^{\pm}_r$, as was already noticed in \cite{deBoer:2008ss}.

To study the fusion rules of two such representations we first collect
what we know based on the fusion rules with respect to the two
$\N=2$ subalgebras (the ones generated by $G^\pm_r$ and
${G'}^{\pm}_r$, respectively). Using the same arguments as in 
the previous section, and combining the constraints coming from the two 
$\N=2$ algebras we know
that the fusion rules of two chiral primary fields are 
\be\label{ccf}
\phi_c \otimes \phi_c = [\varphi_{h=1,q=2}] \oplus
[G^+_{-1/2} {G'}^{+}_{-1/2} \varphi_{h,q=0}] \ .
\ee
Here, the conformal families are written with respect to either of the
$\N=2$ algebras.  
On the other hand, by the arguments of section~3.3 we also know
that the fusion of $\phi_c$ and $\phi_a$ does not allow for any
odd fusion rules, and thus is of the form 
\be\label{caf}
\phi_c \otimes \phi_a = [\tilde\varphi_{h,q=0}] \ .
\ee
Finally, we can use the constraints coming from the affine su(2) 
symmetry. At $k=1$ we know that the fusion of two $j=1/2$ representations
only leads to a $j=0$ representation (since $j=1$ is not allowed at $k=1$). 
Furthermore, if $\tilde\varphi$ is any $\N=4$ highest weight state that appears 
in the fusion of $\phi_c$ with $\phi_a$ it then follows that 
\begin{eqnarray}
0 \neq \langle \tilde\varphi \,| \, \phi_a(1) \, \phi_c(0) \rangle & = &
\langle \tilde\varphi\, | \, (J^-_0\phi_c)(1) \, \phi_c(0) \rangle \nonumber \\
& = &
\langle \tilde\varphi\,  |\,  J^-_1 \Bigl( \phi_c(1) \, \phi_c(0) \Bigr) \rangle \nonumber \\
& = & \langle J^+_{-1} \tilde\varphi\, | \, \phi_c(1)\,  \phi_c(0) \rangle  \ .
\end{eqnarray}
Here we have written out the explicit form of the scalar product and, in the
second line, have made use the comultiplication formula.
Thus for every $\tilde\varphi$  that appears on the right hand side of  
(\ref{caf}), $J^+_{-1}\tilde\varphi$ must appear on the right hand
side of (\ref{ccf})! As we have argued before (based on the
representation theory of the $\N=4$ algebra at $k=1$), the first term
in (\ref{ccf}) is of this form, since 
\be\label{1st}
\varphi_{h=1,q=2} = J^+_{-1} \Omega \ . 
\ee
The second type of terms in (\ref{ccf}) is actually also of this form,
since at $k=1$  we always have the null relation
\be
\Bigl( J^+_{-1} - \frac{1}{2h} G^+_{-1/2} {G'}^{+}_{-1/2} \Bigr) 
\varphi_{h,q=0}= 0 \ ,
\ee
where $\varphi_{h,q=0}$ is in the singlet representation of the su(2)
zero modes, and we have assumed that $h>0$. This shows
that we can write the above fusion rules more compactly as 
\be
\phi_c \otimes \phi_c  =  [J^+_{-1} \varphi_{h,q=0}] \ , \qquad
\phi_c \otimes \phi_a = [\varphi_{h,q=0}] \ ,
\ee
where $\varphi_{h,q=0}$ is in the singlet representation of $\N=4$ and
both instances refer to the {\em same} representation. In particular,
the representations on the right hand side then combine into one
$\N=4$ representation, as must be the case (since the  states on the
left-hand sides also lie in the same $\N=4$ representations). Thus we
can write more succinctly
\be\label{B11}
[\phi_{h=1/2,j=1/2}] \otimes [\phi_{h=1/2,j=1/2}] = [\varphi_{h,j=0}]\,,
\ee
where now the conformal families refer to $\N=4$
families. Unfortunately, we have not been able to deduce any
non-trivial constraints on the possible values of $h$ that  appear on
the right hand side, and the example of section~4.2 suggests that  no
such general constraint exists.

\end{document}